\documentclass[twocolumn,prd,amsmath,amssymb, aps,superscriptaddress, nofootinbib]{revtex4-1}
\usepackage[version=4]{mhchem} 
\usepackage[normalem]{ulem}
\usepackage{amsmath}
\usepackage{amssymb}
\usepackage{amsfonts}
\usepackage{mathrsfs}
\usepackage{graphicx}
\usepackage{xcolor}
\usepackage{xfrac}
\usepackage{footmisc}
\usepackage{pifont}
\usepackage{times}
\usepackage{epstopdf}
\usepackage{enumerate}
\usepackage[hidelinks]{hyperref}
\usepackage{enumitem}
\usepackage{slashed}
\usepackage[absolute,overlay]{textpos} 
\usepackage{relsize}
\usepackage{epsfig}
\usepackage{verbatim}
\usepackage{bm}
\usepackage[utf8]{inputenc}
\usepackage{graphics}
\usepackage{subfigure}
\usepackage[english]{babel}
\usepackage{orcidlink}

\definecolor{zima_blue}{HTML}{1393C1}
\hypersetup{setpagesize=false,
bookmarksnumbered=true,
bookmarksopen=true, 
colorlinks=true,
linkcolor=zima_blue,
urlcolor=zima_blue,
citecolor=zima_blue,
linktocpage=false}

\begin{document}

\begin{textblock}{4}(0.2,0.1)   
  \includegraphics[width=2cm]{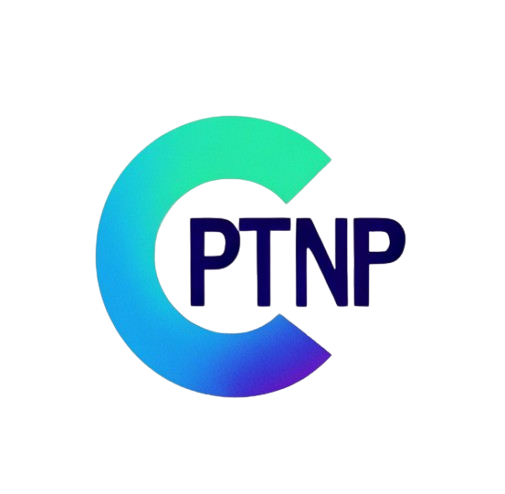}  
\end{textblock}

\begin{textblock}{6}(14,1)
  \raggedleft
  {\text{CPTNP-2025-004}}  
\end{textblock}

\title{Interplay of 95 GeV Diphoton Excess and Dark Matter in Supersymmetric Triplet Model}

\author{Zetian Li}
\affiliation{Department of Physics, Yantai University, Yantai 264005, China}

\author{Ning Liu}
\email{liuning@njnu.edu.cn}
\affiliation{Department of Physics and Institute of Theoretical Physics, Nanjing Normal University, Nanjing, 210023, China}

\author{Bin Zhu}
\email{zhubin@mail.nankai.edu.cn}
\affiliation{Department of Physics, Yantai University, Yantai 264005, China}

\begin{abstract}
The decay of the Higgs boson and the nature of dark matter remain fundamental challenges in particle physics. 
We investigate the $95$~GeV diphoton excess and dark matter within the framework of the triplet-extended Minimal Supersymmetric Standard Model (TMSSM). 
In this model, an additional Hypercharge $Y=0$, $SU(2)_L$ triplet superfield is introduced. 
Mixing between the triplet and doublet Higgs states enhances the diphoton signal strength of the $95$~GeV Higgs boson, resulting in $\mu_{\gamma\gamma}^{\text{CMS+ATLAS}} = 0.24_{-0.08}^{+0.09}$, which is consistent with experimental observations. 
This enhancement arises primarily from charged Higgs and chargino loop contributions. 
Additionally, the model accommodates viable dark matter candidates in the form of a bino-dominated neutralino. 
The relic density is reduced to the observed value through resonance-enhanced annihilation via the Higgs portal or co-annihilation with the triplino or higgsino. 
This reduction remains consistent with constraints from direct and indirect detection experiments. 
A comprehensive parameter scan demonstrates that the TMSSM can simultaneously explain the $95$~GeV diphoton excess, the observed 125~GeV Higgs mass, and the dark matter relic density, establishing a compelling and theoretically consistent framework.
\end{abstract}

\maketitle

\section{Introduction}
\label{s1}

The Standard Model (SM) of particle physics, a cornerstone of our understanding of fundamental interactions, faces significant challenges in explaining the hierarchy problem of Higgs boson mass and the nature of dark matter (DM). The SM cannot stabilize the Higgs mass against quantum corrections, a deficiency known as the hierarchy problem~\cite{Randall:1999ee,Arkani-Hamed:1998jmv}, which results in sensitivity to ultraviolet (UV) divergences. Supersymmetry~\cite{Haber:1984rc,Martin:1997ns} offers a compelling solution by introducing supersymmetric partners that cancel quadratic loop corrections, thereby stabilizing the Higgs mass and reducing fine-tuning. The lightest supersymmetric particle (LSP), stabilized by R-parity, serves as a natural DM candidate, often identified as the neutralino~\cite{Jungman:1995df,Duan:2017ucw}.

The Minimal Supersymmetric Standard Model (MSSM) addresses both the hierarchy problem and DM, but the lack of experimental evidence for superpartners has led to the "little hierarchy problem"~\cite{Kitano:2006gv,Papucci:2011wy,Strumia:2011dv,Craig:2015pha}. This problem arises from the difficulty in reconciling the observed $125$ GeV Higgs boson mass with the non-observation of stops and gluinos~\cite{CMS:2023mny, ATLAS:2024lda}. The Triplet-Extended MSSM (TMSSM)~\cite{Espinosa:1991wt,DiChiara:2008rg,Basak:2012bd,Cort:2013foa,Bandyopadhyay:2013lca,Bandyopadhyay:2014tha,Garcia-Pepin:2014yfa,Bandyopadhyay:2015oga,Bandyopadhyay:2015ifm,Bandyopadhyay:2015tva,Diaz-Cruz:2007rcq,Arina:2014xya,Barradas-Guevara:2004all,Das:2015tva} potentially resolves this challenge by introducing an $SU(2)_L$ triplet superfield. This addition enhances quartic couplings, providing a natural solution to the little hierarchy problem~\cite{FileviezPerez:2011dg}.

Recent data from CMS~\cite{CMS:2018cyk,CMS:2023yay} and ATLAS~\cite{ATLAS:2018xad,ATLAS:2023jzc} report a significant signal for a $95$ GeV Higgs boson decaying to diphotons. CMS observes a local excess with $2.9\sigma$ significance and a signal strength of $\mu_{\gamma \gamma} = 0.33_{-0.12}^{+0.19}$, while ATLAS finds $\mu_{\gamma \gamma} = 0.18 \pm 0.10$. The combined signal strength, $\mu_{\gamma \gamma}^{\text{CMS+ATLAS}} = 0.24_{-0.08}^{+0.09}$~\cite{Biekotter:2023oen}, suggests the 95 GeV Higgs may exceed the SM prediction, indicating potential new physics.

Triplet-extended models introduce an $SU(2)_L$ triplet scalar field, which can explain the $95$~GeV diphoton excess through mixing with the SM Higgs doublet, enhancing the diphoton decay rate. This mechanism has been extensively studied in non-supersymmetric frameworks~\cite{Cao:2016uwt,Crivellin:2017upt,Haisch:2017gql,Fox:2017uwr,Liu:2018xsw,Wang:2018vxp,Liu:2018ryo,Biekotter:2019kde,Cline:2019okt,Choi:2019yrv,Kundu:2019nqo,Cao:2019ofo,Biekotter:2020cjs,Abdelalim:2020xfk,Heinemeyer:2021msz,Biekotter:2022jyr,Iguro:2022dok,Li:2022etb,Biekotter:2023jld,Bonilla:2023wok,Azevedo:2023zkg,Biekotter:2023oen,Escribano:2023hxj,Belyaev:2023xnv,Ashanujjaman:2023etj,Baek:2024cco,Du:2025eop,YaserAyazi:2024hpj,Borah:2023hqw,Bhattacharya:2023lmu,Coloretti:2023yyq,Banik:2023vxa,Ashanujjaman:2024lnr,Dev:2023kzu,Mondal:2024obd}, where precise tuning of triplet-doublet mixing and mass splittings between charged and neutral Higgs states amplifies the diphoton signal strength while satisfying experimental constraints. In supersymmetry, the singlet-extended MSSM has also been thoroughly investigated~\cite{Cao:2023gkc,Ellwanger:2023zjc,Lian:2024smg,Ellwanger:2024vvs,Ellwanger:2024txc,Cao:2024axg,Liu:2024cbr}. Given this promising property, it is compelling to explore whether the TMSSM, the supersymmetric generalization, retains this feature. In the TMSSM, supersymmetry imposes strict relations on coupling constants, avoiding the artificial adjustments often required in non-supersymmetric models. This rigidity may preserve the enhanced diphoton signal necessary to explain the $95$~GeV excess while eliminating the need for fine-tuning couplings, offering a significant advantage over non-supersymmetric approaches.

Additionally, the triplet Higgs field introduces new annihilation channels for neutralinos~\cite{Arina:2014xya}, especially when the triplet mass is approximately twice the neutralino mass, around $95$ GeV or $125$ GeV. The triplino, the supersymmetric partner of the triplet Higgs, plays a crucial role in neutralino annihilation. We demonstrate that even if the triplino is not the LSP, it can significantly influence DM annihilation by modifying chargino masses. This modification substantially alters the DM relic density through bino-triplino co-annihilation processes~\cite{Edsjo:1997bg}. 

In this work, we show that the TMSSM simultaneously accounts for the $95$~GeV diphoton excess and dark matter phenomenology. Through a detailed analysis of the Higgs and neutralino sectors and a comprehensive parameter scan, we identify viable parameter regions that satisfy all relevant constraints, presenting a compelling extension of the SM.

This paper is structured as follows: Section~\ref{s2} presents the TMSSM framework, focusing on the impact of the $SU(2)_L$ triplet superfield on the Higgs sector. We analytically demonstrate how the triplet accounts for the 95~GeV Higgs and its diphoton excess. Section~\ref{s3} explores the roles of the triplet and triplino in dark matter phenomenology, analyzing their effects on annihilation cross-sections and relic density while ensuring consistency with detection constraints. Section~\ref{s4} evaluates the TMSSM's ability to simultaneously address the Higgs decay and dark matter problems through a comprehensive parameter scan. Section~\ref{s5} summarizes the findings and discusses their broader implications for physics beyond the Standard Model.

\section{Higgs sector: Higgs mass and its decay}
\label{s2} 

The TMSSM extends the MSSM through the introduction of an additional $SU(2)_L$-triplet superfield $\Sigma$, 
\begin{equation}
\Sigma = \begin{pmatrix} 
T^{0}/\sqrt{2} & T^{+} \\ 
T^{-} & -T^{0}/\sqrt{2} 
\end{pmatrix}.
\end{equation}
The TMSSM superpotential adds two terms: a triplino mass term $\mu_T$ and a cubic coupling $\lambda$ that mixes the triplet with the Higgs doublets.  The full superpotential reads,
\begin{equation}
W_{\mathrm{TMSSM}}=W_{\mathrm{MSSM}}+\lambda H_u \cdot \Sigma H_d+\mu_{T} \operatorname{Tr} \Sigma^2.
\end{equation}

The model also incorporates soft SUSY-breaking terms:
\begin{equation}
\begin{aligned}
    \mathcal{L}_{\mathrm{soft}} &=\mathcal{L}_{\mathrm{MSSM}_{\mathrm{SB}}}+m_T^2 \operatorname{Tr}\left(\Sigma^{\dagger} \Sigma\right)+B_{\mu_T} \operatorname{Tr}\left(\Sigma^2\right)\\
    &+T_\lambda H_u \cdot \Sigma H_d+\text { h.c. },
    \end{aligned}
\end{equation}
where only triplet-specific terms are explicitly listed. To simplify the analysis, we choose a SUSY-breaking scale $M_S$ to unify the soft masses $M_S=m_{\tilde{Q}} = m_{\tilde{u}} = m_{\tilde{d}} = m_{\tilde{L}} = m_{\tilde{e}}$, and $A_0$ to represent the trilinear soft terms $A_0 = A_u = A_d = A_e$. The remaining soft masses $m_{H_u}^2$, $m_{H_d}^2$, and $m_T^2$ are determined by imposing the conditions for successful electroweak symmetry breaking (EWSB), i.e., the tadpole equations $\partial V / \partial \phi_i = 0$, where $i$ corresponds to $H_u$, $H_d$, and $T$. 

During the EWSB process, the neutral component $T^{0}$ of the triplet acquires a vacuum expectation value (VEV). The neutral components of the Higgs doublets and the triplet can be expressed as fluctuations around their respective VEVs:
\begin{equation}
\begin{aligned}
H_d^0 &= \frac{1}{\sqrt{2}} \left( v_d + \phi_d + i \sigma_d \right),\\
H_u^0 &= \frac{1}{\sqrt{2}} \left( v_u + \phi_u + i \sigma_u \right),\\
T^0 &= \frac{1}{\sqrt{2}} \left( v_T + \phi_T + i \sigma_T \right),
\end{aligned}
\end{equation}
where $v_{d}$, $v_{u}$, and $v_{T}$ are the VEVs of the Higgs doublets and the triplet, respectively. The terms $\phi_d$, $\phi_u$, and $\phi_T$ represent their real parts, while $\sigma_d$, $\sigma_u$, and $\sigma_T$ denote their imaginary parts. 

The electroweak scale is defined by the Higgs vacuum expectation values (VEVs) as $v = \sqrt{v_d^2 + v_u^2} = 246~\mathrm{GeV}$. Triplet-extended models face constraints from electroweak precision tests (EWPT), particularly through the $\rho$ parameter. A non-zero triplet VEV $v_T$ alters $\rho$ from unity according to $\rho = 1 + \frac{4v_T^2}{v^2}$. With the experimental value $\rho = 1.0004^{+0.0003}_{-0.0004}$, this limits $v_T \lesssim 4~\mathrm{GeV}$. However, introducing two additional triplet chiral superfields with hypercharges $Y = \pm 1$, alongside the $Y = 0$ triplet, preserves custodial symmetry in the superpotential~\cite{Delgado:2015aha}. This ensures $\rho = 1$ at tree level, permitting a larger triplet VEV $v_T$ while remaining consistent with EWPT constraints.

Due to the mixing between different gauge eigenstates, $\left( \phi_{d}, \phi_{u}, \phi_{T} \right)$, they form the non-diagonal CP-even Higgs mass matrix. 

\begin{equation}
    \mathcal{M}_{h}^{2} = 
    \scalebox{1.2}{
        $\begin{pmatrix}
            \mathcal{M}_{\phi_{d}\phi_{d}}  & \mathcal{M}_{\phi_{u}\phi_{d}} & \mathcal{M}_{\phi_{T}\phi_{d}} \\
            \mathcal{M}_{\phi_{d}\phi_{u}}  & \mathcal{M}_{\phi_{u}\phi_{u}} & \mathcal{M}_{\phi_{T}\phi_{u}} \\
            \mathcal{M}_{\phi_{d}\phi_{T}}  & \mathcal{M}_{\phi_{u}\phi_{T}} & \mathcal{M}_{\phi_{T}\phi_{T}}
        \end{pmatrix}$
    }.
    \label{eqn:massmatrix}
\end{equation}

In the limit of small $v_T$ and $\lambda$, the triplet sector almost decouples from the doublet sector. Consequently, the 95 GeV Higgs predominantly originates from the triplet itself. We therefore analyze the $(33)$-th component of the mass matrix first and denote the lightest higgs mass by $m_h$,

\begin{equation} 
m_h^2=\mathcal{M}_{\phi_T\phi_T} = 2 B_{\mu_T}  + 4 \mu_T^2 + m_T^2,
\label{eqn:Triplet}
\end{equation}

where $m_T^2$ can be determined by solving the tadpole equations at the EWSB scale $\partial  V/\partial \phi_T=0$,

\begin{equation}
\begin{aligned}
    m_T^2 =& -2B_{\mu_T}-4\mu_T^2 {+} \frac{\lambda}{2}\left(\frac{v}{v_T}\right) v\mu_T\sin(2\beta)\\
    &{-}\frac{\lambda}{4}\left(\frac{v}{v_T}\right) v\mu \sin^2\beta,
    \label{eqn:mT2}
\end{aligned}
\end{equation}

This relation holds exclusively at tree-level, neglecting terms linear in $\lambda$ and $v_T$. We note, however, that a linear term in $\lambda$ persists in Eq.~\eqref{eqn:mT2} due to the $v/v_T$ enhancement, which counteracts the $\lambda$ suppression. Substituting $m_T^2$ from Eq.~\eqref{eqn:mT2} into the triplet Higgs mass (Eq.~\eqref{eqn:Triplet}), we observe a precise cancellation of $B_{\mu_T}$,

\begin{equation}
    m_{h}^2= \frac{\lambda}{2}\left(\frac{v}{v_T}\right) v\mu_T\sin(2\beta)-\frac{\lambda}{4}\left(\frac{v}{v_T}\right) v\mu \sin^2\beta.
    \label{eqn:expression}
\end{equation}

\begin{figure*}
\includegraphics[width=0.49\textwidth]{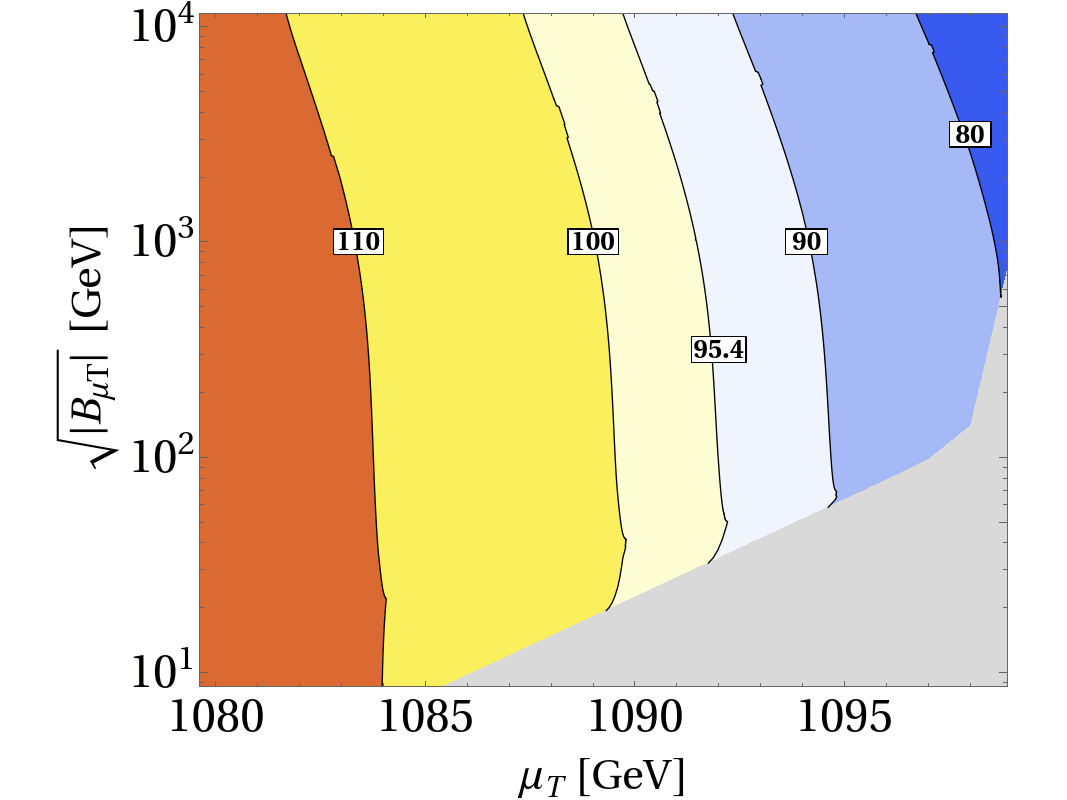}
\includegraphics[width=0.49\textwidth]{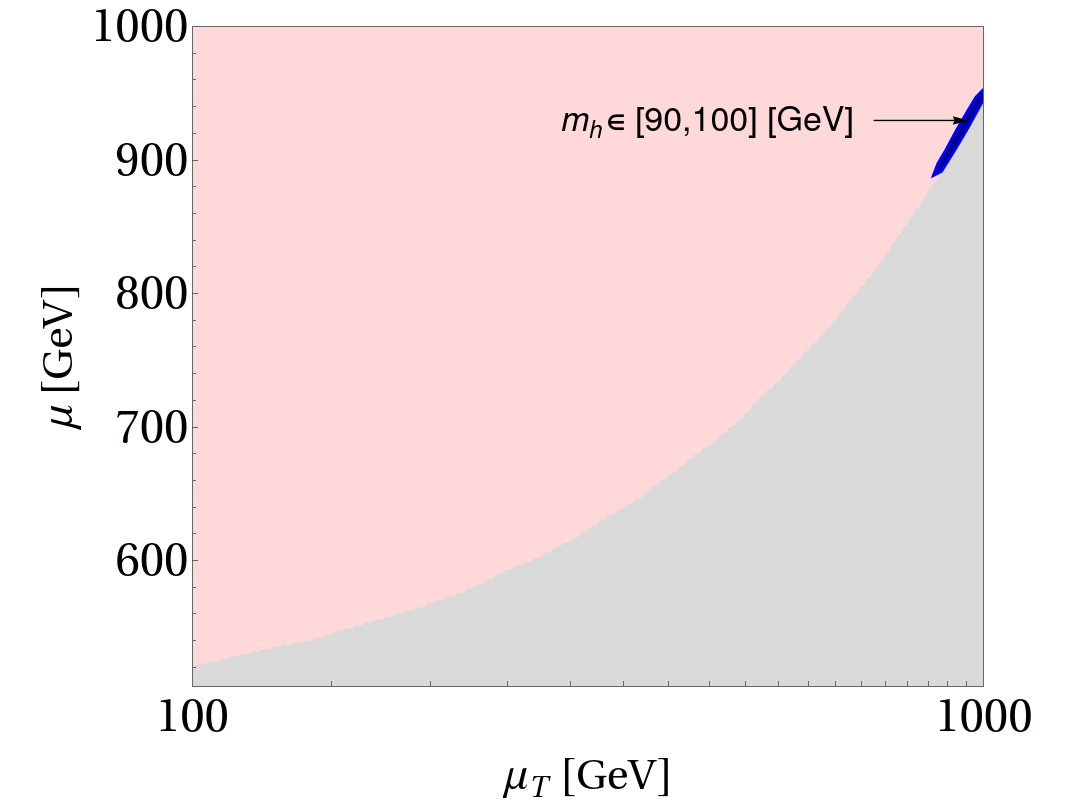}
\caption{Left: Contour plot of the lightest Higgs mass $m_h$ in the $\sqrt{B_{\mu_T}} -\mu_T$ parameter space, with $\mu=1~\mathrm {TeV}$, $M_S=5.01~\mathrm {TeV}$, $A_0=12~\mathrm {TeV}$.
Right: Contour plot of the lightest Higgs boson mass $m_h$ in the $\mu-\mu_T$ parameter space, with $M_S=5.01~\mathrm {TeV}$, $A_0=12~\mathrm {TeV}$, and $\sqrt{B_{\mu_T}}  =\sqrt{5}\mathrm{TeV}$.}
\label{fig:Higgs95.4}
\end{figure*}

This cancellation explains the independence of the lightest Higgs boson mass on $B_{\mu_T}$ in Fig.~\ref{fig:Higgs95.4} (left panel). The residual minor dependence on $B_{\mu_T}$ arises from loop-level corrections. 

The right panel of Fig.~\ref{fig:Higgs95.4} shows that the mass of the triplet-dominated lightest Higgs is inversely proportional to $\mu_T$, which appears to contradict the linear dependence on $\mu_T$ predicted by Eq.~\eqref{eqn:expression}. This apparent contradiction is resolved by the negative cubic coupling $\lambda$, which reverses the proportionality. Consequently, the lightest Higgs mass scales inversely with $\mu_T$, as depicted in Fig.~\ref{fig:Higgs95.4}. We numerically exclude the light-gray region where the squared mass of the charged Higgs, $m_{H^\pm}^2$, is negative. 

\begin{figure*}
\includegraphics[width=0.49\textwidth]{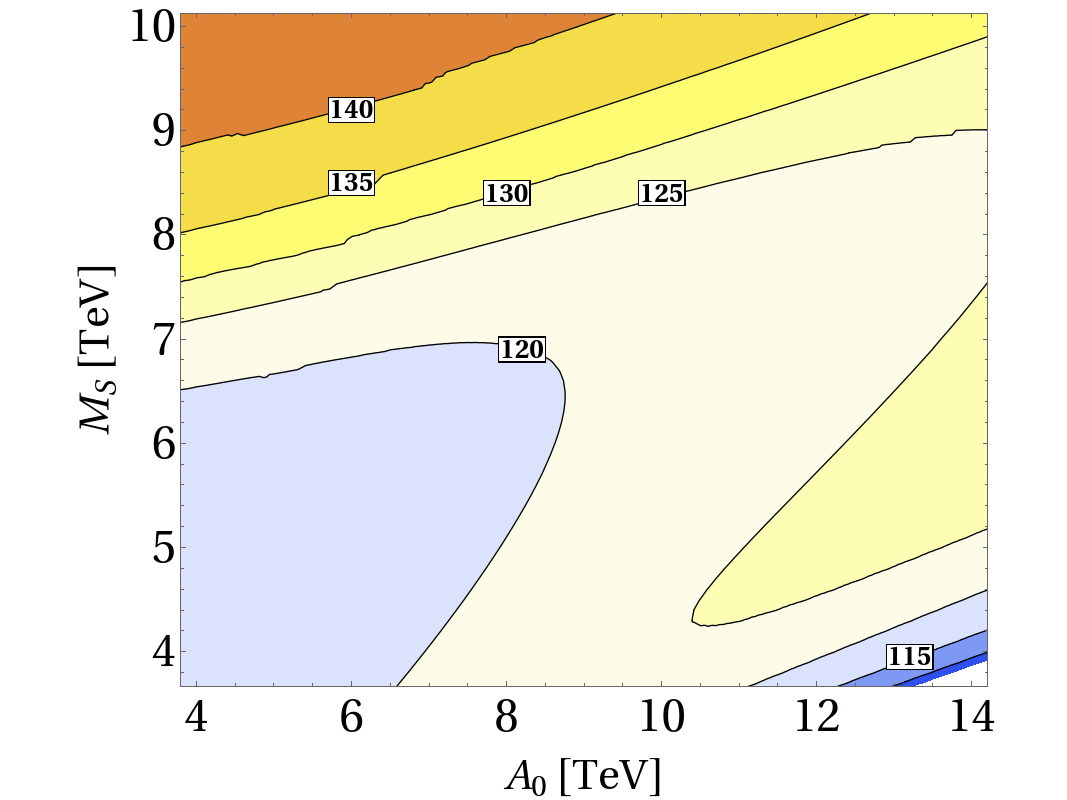}
\includegraphics[width=0.49\textwidth]{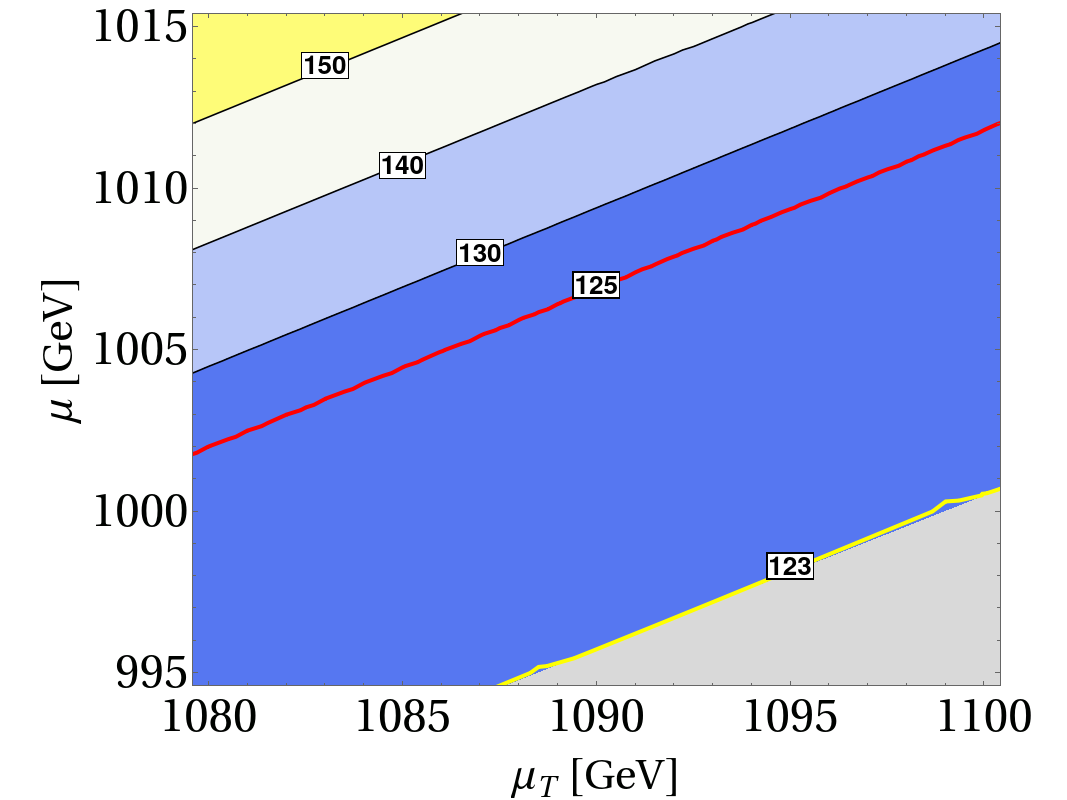}
\caption{
Left: Contour plot of the SM Higgs mass $m_H$ in the $A_0$-$M_S$ parameter space, with $\mu=1~\mathrm {TeV}$, $\mu_T=1.092~\mathrm {TeV}$, and $B_{\mu_T}=5\left ( \mathrm {TeV}\right ) ^{2}  $. 
Right: Contour plot of the SM Higgs boson mass $m_H$ in the $\mu-\mu_T$ parameter space, with $M_S=5.01~\mathrm {TeV}$, $A_0=12~\mathrm {TeV}$, and $B_{\mu_T}    =5\left ( \mathrm {TeV}\right ) ^{2} $.} 
\label{fig:Higgs125}
\end{figure*}

The doublet Higgs is responsible for the observed $125~\mathrm{GeV}$ Higgs boson. The tree-level Higgs mass enhancement, proportional to $\lambda^2 v^2 \sin^2 2\beta$, is suppressed due to the small value of $\lambda$. Therefore, we rely on a high supersymmetry (SUSY) scale and large trilinear couplings to achieve the required Higgs mass. The left panel of Fig.~\ref{fig:Higgs125} illustrates that maximal mixing ($A_0 \sim M_S$) greatly enhances the Higgs mass, avoiding the need for very large $M_S$ values. When $A_0$ is small, the Higgs mass depends more on the SUSY scale $M_S$, requiring larger $M_S$ for higher Higgs boson masses.

The right panel of Fig.~\ref{fig:Higgs125} reveals that the triplet parameter $\mu_T$ influences the SM Higgs boson mass, despite the nominal decoupling of the triplet sector. Direct inspection shows that the $\mathcal{M}_{\phi_d\phi_d}$ and $\mathcal{M}_{\phi_u\phi_u}$ mass matrix components depend on $\mu_T$ through the highly suppressed term $\lambda v_T/v$, allowing these contributions to be safely neglected. The $\mu_T$ dependence instead arises from the off-diagonal mixing terms $\mathcal{M}_{\phi_u\phi_T}$ and $\mathcal{M}_{\phi_d\phi_T}$. For example, the $\mathcal{M}_{\phi_u\phi_T}$ component is

\begin{equation}
\mathcal{M}_{\phi_u \phi_T} = -\lambda v( \mu_T  \cos\beta -\mu\sin\beta) + \mathcal{O}(\lambda^2).
\end{equation}

Despite $\lambda \sim \mathcal{O}(-0.1)$,  its residual contribution induces a measurable shift in the$95$ and $125$ GeV Higgs boson mass. The mixing between the doublet and triplet Higgs fields $h= \phi_{\mathrm{SM}}\cos\theta+ \phi_T\sin\theta$ is essential for the production of the lightest neutral Higgs boson via gluon fusion. Without this mixing, gluon fusion—the dominant production mechanism—would be absent, as the triplet Higgs does not couple directly to fermions.

Alternative processes, such as vector boson fusion or associated production, typically yield insufficient cross-sections. Thus, doublet-triplet mixing is necessary to achieve a sufficiently large production cross-section via gluon fusion, explaining the observed diphoton excess, which is determined as follows,

\begin{equation}
    \begin{aligned}
\mu_{\gamma \gamma } =\frac{\sigma _{\mathrm{SUSY}}\left ( pp\to h \right )\times \mathrm{Br}_{\mathrm{SUSY}}\left ( h\to \gamma \gamma  \right )    }{\sigma _{\mathrm{SM}}\left ( pp\to h \right )\times \mathrm{Br}_{\mathrm{SM}}\left ( h\to \gamma \gamma  \right )}.
\end{aligned}
\end{equation}

To influence $\mu_{\gamma\gamma}$, modifications to either the Higgs production mechanism or its decay width are required. Owing to the small doublet-triplet mixing angle $\theta$, the gluon fusion production cross-section remains significant, analogous to the SM mechanism. The gluon fusion cross-section for $h$ is,
\begin{equation}
\sigma_{ggF} \propto \left| \cos\theta \cdot A_t(\tau_t) + A_{\tilde{t}}(\tau_{\tilde{t}}) \right|^2,
\end{equation}
where $A_t(\tau_t) = \frac{3}{2} \tau_t \left[ 1 + (1 - \tau_t) f(\tau_t) \right]$ (top quark loop) and $A_{\tilde{t}}$ (stop loop) are loop functions with $\tau_i = 4m_i^2/m_h^2$. The SM cross-section is $\sigma_{ggF}^{\text{SM}} \propto |A_t^{\text{SM}}|^2$, yielding the ratio,
\begin{equation}
\frac{\sigma_{\mathrm{SUSY}}}{\sigma_{\mathrm{SM}}} =\frac{\sigma_{ggF}}{\sigma_{ggF}^{\text{SM}}} = \left| \cos\theta + \frac{A_{\tilde{t}}}{A_t^{\text{SM}}} \right|^2\sim 1.
\end{equation}

For small mixing and large SUSY scale $M_S$, the production cross-section ratio approaches the Standard Model. Thus, deviations in the diphoton signal arise predominantly from the decay width~\cite{Djouadi:2005gj},
\begin{widetext}
\begin{equation}
\Gamma_{h\rightarrow\gamma\gamma}=\frac{\alpha^2 m_h^3}{1024 \pi^3}\left|\frac{g_{h V V}}{m_V^2} Q_V^2 A_1\left(\tau_W\right)+\frac{2 g_{h f \bar{f}}}{m_f} N_{c, f} Q_f^2 A_{1 / 2}\left(\tau_f\right)+N_{c, S} Q_S^2 \frac{g_{h S S}}{m_S^2} A_0\left(\tau_S\right)\right|^2.
\end{equation}
\end{widetext}
with loop amplitudes:
\begin{itemize}
\item $A_1(\tau_W)$ is the loop function of $W$-boson.
\item $\left.A_0\left(\tau_s\right)=-\tau_i^2\left[\tau_s^{-1}-f\left(\tau_s^{-1}\right)\right]\right]$ is the loop function of triplet charged Higgs.
\item $A_{1/2}(\tau_f)$ is the loop function of chargino loop.
\end{itemize}

They reveal that additional contributions from charged Higgs bosons and charginos modify the Higgs decay width. For charginos, although heavier masses would naively induce larger deviations, their contribution is suppressed by the loop function $A_{1/2}(\tau_{\chi^\pm})$. This interplay naturally prevents unphysical growth of deviations while maintaining consistency with the decoupling theorem.  Thus, the dominant contribution arises from the lightest charged Higgs boson (charged triplet) and moderate chargino mass, as numerically verified in Section~\ref{s4}.

\section{Neutralino sector: Relic Density and Direct Detection} 
\label{s3}

The neutralino mass matrix includes the traditional neutralino components: the Bino ($\lambda_{\tilde{B}}$), Wino ($\tilde{W}^0$), and Higgsinos ($\tilde{H}_d^0$, $\tilde{H}_u^0$), as well as the newly introduced fermionic triplet component, the triplino ($\lambda_{T^0}$). The specific form of the mass matrix is given as follows,

\begin{equation}
    \begin{pmatrix} M_{1}  & 0 & -\frac{1}{2}g_{1}v_{d}    & \frac{1}{2}g_{1}v_{u} & 0\\  0 & M_{2} & \frac{1}{2}g_{2}v_{d} & -\frac{1}{2}g_{2}v_{u} & 0\\ -\frac{1}{2}g_{1}v_{d} & -\frac{1}{2}g_{2}v_{d} & 0 & -\frac{1}{2}v_{T}\lambda -\mu    & -\frac{1}{2}v_{u}\lambda \\ \frac{1}{2}g_{1}v_{u} &  -\frac{1}{2}g_{2}v_{u}& -\frac{1}{2}v_{T}\lambda -\mu & 0 &-\frac{1}{2}v_{d}\lambda \\ 0 & 0 & -\frac{1}{2}v_{u}\lambda  &  -\frac{1}{2}v_{d}\lambda & 2\mu_{T}\end{pmatrix}.
\end{equation}

To investigate the influence of various components on the DM annihilation process,  the diagonal process can be employed to study the contribution of the triplino to DM in the TMSSM. First, it is essential to diagonalize the neutralino mass matrix to derive its eigenvalue mass matrix,

\begin{equation}
    N^{\ast } m_{\tilde{\chi}_{0}   } N^{\dagger } =m_{\tilde{\chi}_{0}  }^{\text{dia}}. 
\end{equation}

The contributions based on the components in the neutralino basis can be expressed as,

\begin{equation}
\begin{aligned}
\lambda _{\tilde{B} } &=\sum_{j}N_{j1}^{\ast } \lambda _{j}^{0} ,\quad 
\tilde{W } ^{0} =\sum_{j}N_{j2}^{\ast } \lambda _{j}^{0} ,\quad 
\tilde{H}_{d}^{0}  =\sum_{j}N_{j3}^{\ast } \lambda _{j}^{0} ,\\
\tilde{H}_{u}^{0}  &=\sum_{j}N_{j4}^{\ast } \lambda _{j}^{0} ,\quad 
\lambda _{T^{0} }=\sum_{j}N_{j5}^{\ast } \lambda _{j}^{0}.
\end{aligned} 
\end{equation}

We select two benchmark points to examine their contributions to dark matter (DM) composition, as presented in Table~\ref{tab:ZN_matrix}. These points represent Bino-dominated and Triplino-dominated DM scenarios. We exclude explicit discussion of the well-known Higgsino and Wino LSP scenarios.
 
In the parameter selection for ID A, the values are set as follows: $M_1 = 62.5~\mathrm{GeV}\ll M_2, \mu, \mu_T$, and $\lambda = -0.1$. Specifically, the choice of $M_1$ aligns the bino mass precisely with the Higgs resonance region, significantly enhancing the annihilation efficiency via $s$-channel Higgs exchange. 

\begin{figure}
    \centering
    \includegraphics[width=1\linewidth]{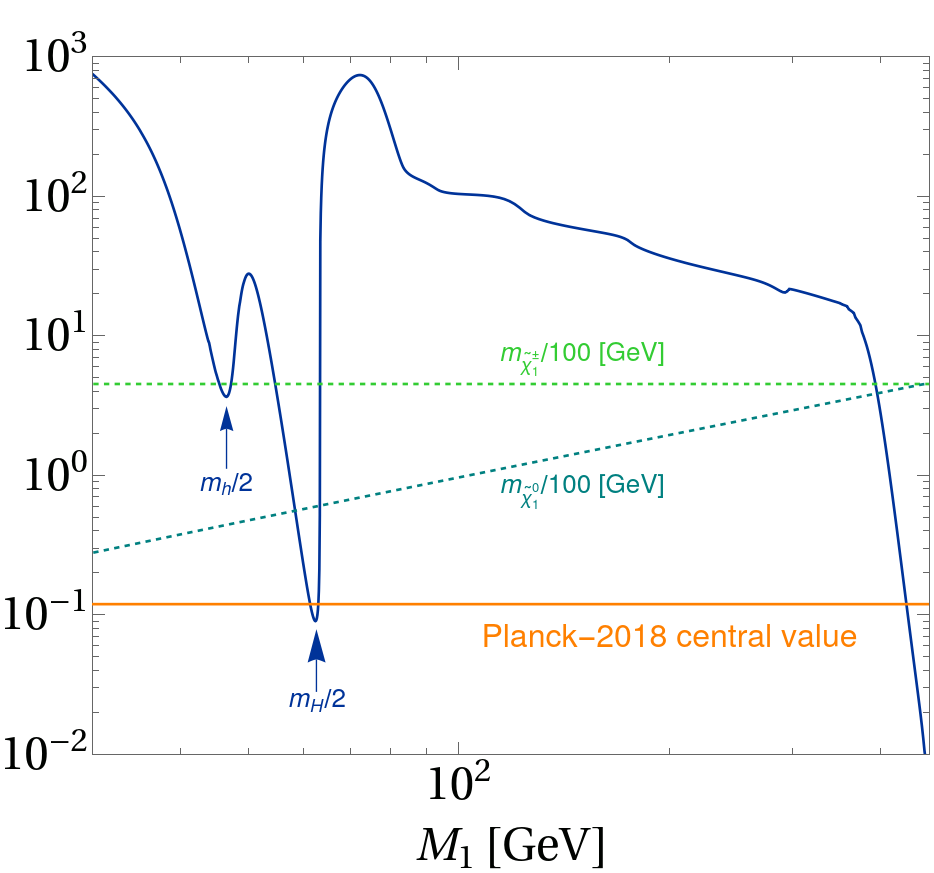}
    \caption{The blue solid line is the neutralino relic density with the variation of $M_1$. The orange solid line is used to highlight the central relic density value from the Planck 2018 data. The green dashed line and cyan dashed line correspond to lightest chargino and lightest neutralino mass respectively.
    The input parameters are set as follows:~$\tan \beta = 3.95,M_S= 5\,\text{TeV},\lambda= -0.133,\mu_T= 200\,\text{GeV},\mu= 1\,\text{TeV}, B_{\mu_T}=5\left( \text{TeV} \right)^2$. }
    \label{fig:DM1}
\end{figure}

In Fig.~\ref{fig:DM1}, the blue curve depicts the neutralino relic density, and the orange line indicates the observed dark matter relic density, $\Omega h^2 = 0.12$. The green dashed line shows the lightest neutralino mass, which increases linearly with $M_1$, confirming its dominant bino composition. The bino relic density typically exceeds $\Omega h^2 = 0.12$, but two resonant regimes---where $2m_{\tilde{B}} \approx m_h$ or $2m_{\tilde{B}} \approx m_H$---enable resonant annihilation, sharply reducing the relic density. Additionally, when the cyan dashed line (chargino mass) nears the bino mass, co-annihilation processes further decrease the relic density.

Therefore, analyzing the neutralino mixing matrix $Z_N$ alone is insufficient to fully characterize the triplet sector's role in DM physics. As evidenced in Fig.~\ref{fig:DM1}, Bino–Chargino co-annihilation processes significantly enhance the DM annihilation cross-section~\cite{Abdughani:2017dqs,Duan:2018rls,Abdughani:2019wss,Duan:2018cgb}. A complete analysis therefore requires inclusion of the chargino mass matrix, defined in the basis $\left( \tilde{W}^-, \tilde{H}_d^-, \lambda_{T^-} \right)$ and $\left( \tilde{W}^+, \tilde{H}_u^+, \lambda_{T^+} \right)$. The chargino mass matrix is,

\begin{equation}
    m_{\tilde{\chi }^{-}  }= \begin{pmatrix} M_{2}  & \frac{1}{\sqrt{2} }g_{2} v_{u}   &-g_{2}v_{T}   \\ \frac{1}{\sqrt{2} }g_{2} v_{d} & -\frac{1}{2 }v_{T} \lambda +\mu & -\frac{1}{\sqrt{2} }v_{u}\lambda \\ g_{2}v_{T}   & \frac{1}{\sqrt{2}  }v_{d}\lambda  &2\mu_{T}\end{pmatrix}.
\end{equation}

In the limit of small $\lambda$ and $v_T$, the mass matrix becomes nearly diagonal, showing that $\mu_T$ governs the charged triplino mass. This parameter also controls the masses of the lightest Higgs boson and triplino, allowing simultaneous study of Higgs and dark matter phenomenology through $\mu_T$. 

For the bino annihilation, the dominant process in the Higgs resonance region is bino pairs annihilating into bottom quarks via $s$-channel exchange of the Higgs boson~\cite{Arkani-Hamed:2006wnf}. This is natural, as the third-generation Yukawa coupling is the most significant. Additionally, Bino–Chargino co-annihilation processes happens naturally when $\mu_T \approx M_1$, with the dominant channels being $\tilde{\chi}_1^0 + \tilde{\chi}_1^\pm \to Z + W^{-}$ and $\tilde{\chi}_3^0 + \tilde{\chi}_1^\pm \to Z + W^{-}$.

\begin{figure}
    \centering
    \includegraphics[width=1\linewidth]{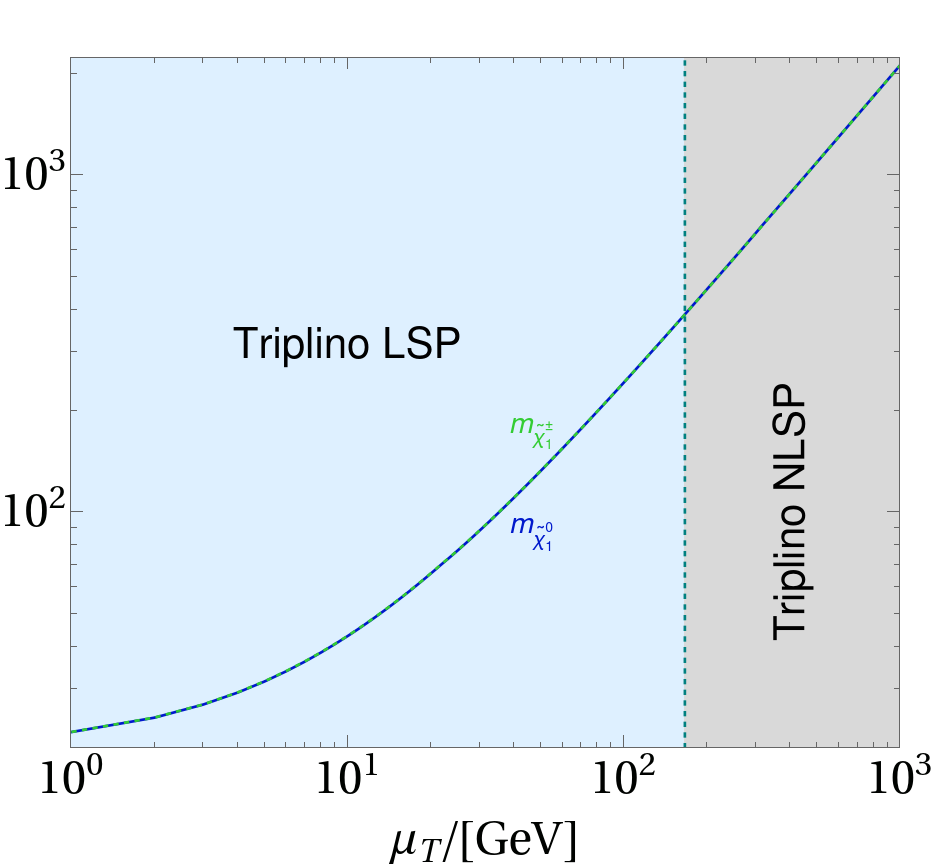}
    \caption{ The masses of the charged triplino and neutral triplino as a function of $\mu_T$, and the status for the LSP.}
    \label{fig:LSP}
\end{figure}

In parameter set ID B, we set $\mu_T = 50~\mathrm{GeV}$ with $\mu_T \ll M_1, M_2, \mu$, ensuring the triplino is the LSP. However, the relic density is low, approximately $10^{-4}$, due to strong $\mathrm{SU}(2)$ interactions similar to those of the wino, which requires a mass of $\sim 2~\mathrm{TeV}$ to match the observed relic density. Although a heavy triplino LSP appears viable for dark matter, Fig.~\ref{fig:LSP} shows that the triplino ceases to be the LSP above $\sim 100~\mathrm{GeV}$. At tree level, with $\mu \gg M_1, M_2$, the neutral and charged triplinos are nearly degenerate ($m \approx 2\mu_T$) and dominate the LSP, as shown in Fig.~\ref{fig:LSP}. As $\mu_T$ increases, loop corrections raise the neutral triplino’s mass more than the charged triplino’s, because the neutral triplino, a Majorana fermion, experiences larger $Z$-boson and scalar loop effects, while the charged triplino, a Dirac fermion, incurs smaller photon and $W$-boson corrections. This inverts the mass hierarchy, making the charged triplino the LSP. Consequently, we adopt a bino LSP for this study.

\begin{table*}[htbp]
\centering
\setlength{\tabcolsep}{8pt} 
\renewcommand{\arraystretch}{1.5} 
\begin{tabular}{l|ccccc|c}
\hline
\multicolumn{7}{c}{$Z^{N}$ and $Z^{H}$ Matrices} \\ \hline
ID A & $Z^{N}(1,1)$ & $Z^{N}(1,2)$ & $Z^{N}(1,3)$ & $Z^{N}(1,4)$ & & \\
     & $-9.9918 \times 10^{-1}$ & $5.1812 \times 10^{-4}$ & $-3.8790 \times 10^{-2}$ & $1.1637 \times 10^{-2}$ & & $\Omega h^{2}$ \\
     & $Z^{N}(1,5)$ & $Z^{H}(1,1)$ & $Z^{H}(1,2)$ & $Z^{H}(1,3)$ & & $0.115$ \\
     & $1.6007 \times 10^{-4}$ & $3.4175 \times 10^{-3}$ & $6.0784 \times 10^{-2}$ & $9.9815 \times 10^{-1}$ & & \\ \hline
ID B & $Z^{N}(1,1)$ & $Z^{N}(1,2)$ & $Z^{N}(1,3)$ & $Z^{N}(1,4)$ & & \\
     & $-2.0387 \times 10^{-4}$ & $4.7161 \times 10^{-4}$ & $-2.2499 \times 10^{-2}$ & $1.3659 \times 10^{-2}$ & & $\Omega h^{2}$ \\
     & $Z^{N}(1,5)$ & $Z^{H}(1,1)$ & $Z^{H}(1,2)$ & $Z^{H}(1,3)$ & & $5.8860 \times 10^{-4}$ \\
     & $9.9991 \times 10^{-1}$ & $1.3531 \times 10^{-2}$ & $6.5181 \times 10^{-2}$ & $9.9778 \times 10^{-1}$ & & \\ \hline
\end{tabular}
\caption{Composition of Bino LSP and Triplino as Dark Matter and their $Z^{H}$.}
\label{tab:ZN_matrix}
\end{table*}

We now consider the direct detection of neutralino dark matter~\cite{Yang:2024uoq}, with a dominant bino component as established~\cite{Cheung:2012qy}. The spin-independent scattering cross-section between the neutralino and a nucleus originates from the effective Lagrangian~\cite{Basak:2013eba},
\begin{equation}
\mathcal{L}_{\text{eff}} = a_q \bar{{\chi}}_1^0 {\chi}_1^0 \bar{q} q,
\end{equation}
where $a_q$ parametrizes the neutralino-quark coupling, and the resulting neutralino-nucleon cross-section becomes
\begin{equation}
\sigma_{\text{SI}} = \frac{4 \mu_r^2}{\pi} \left(\sum_{q=u,d,s} f_{Tq}^{(p,n)} a_q \frac{m_{p,n}}{m_q} + \frac{2}{27} f_{TG}^{(p,n)} \sum_{q=c,b,t} a_q \frac{m_{p,n}}{m_q}\right)^2,
\end{equation}
with $\mu_r$ the neutralino-nucleon reduced mass and $f_{TG}^{(p,n)} = 1 - \sum_{q=u,d,s} f_{Tq}^{(p,n)}$ and form factors are $f_{Tu}^{(p)}=0.020 \pm 0.004$, $f_{Td}^{(p)}=0.026 \pm 0.005$, $f_{Ts}^{(p)}=0.118 \pm 0.062$, $f_{Tu}^{(n)}=0.014 \pm 0.003$, $f_{Td}^{(n)}=0.036 \pm 0.008$, $f_{Ts}^{(n)}=0.118 \pm 0.062$~\cite{Ellis:2000ds}. The dominant contribution to $a_q/m_q$ stems from $t$-channel CP-even Higgs exchange, scaling as
\begin{equation}
\frac{a_q}{m_q} \simeq \sum_i \frac{S_{\chi \chi h_i}}{m_{h_i}^2} S_{h_i qq}.
\end{equation}
For the lightest Higgs with mass $m_{h} \simeq 95.4~\text{GeV}$, the coupling is
\begin{equation}
\begin{aligned}
S_{\chi \chi h_1} \simeq\ & g_2 (Z^N_{12} - \tan\theta_W Z^N_{11})(Z^H_{11} Z^N_{13} - Z^H_{12} Z^N_{14}) \\
& + \lambda ( Z^H_{11} Z^N_{15} Z^N_{13} + Z^H_{12} Z^N_{15} Z^N_{14} + Z^H_{13} Z^N_{13} Z^N_{14} ) 
\end{aligned}
\end{equation}
where $Z^N_{ij}$, $Z^H_{ij}$, and $\lambda$ denote neutralino mixing, Higgs mixing, and model parameters, respectively.Therefore, we obtain the cross section for ID A as $1.0110 \times 10^{-11}$ pb, and for ID B as $6.3827 \times 10^{-15}$ pb.

\section{Numerical Result}
\label{s4}

\begin{figure*}
\includegraphics[width=0.49\textwidth]{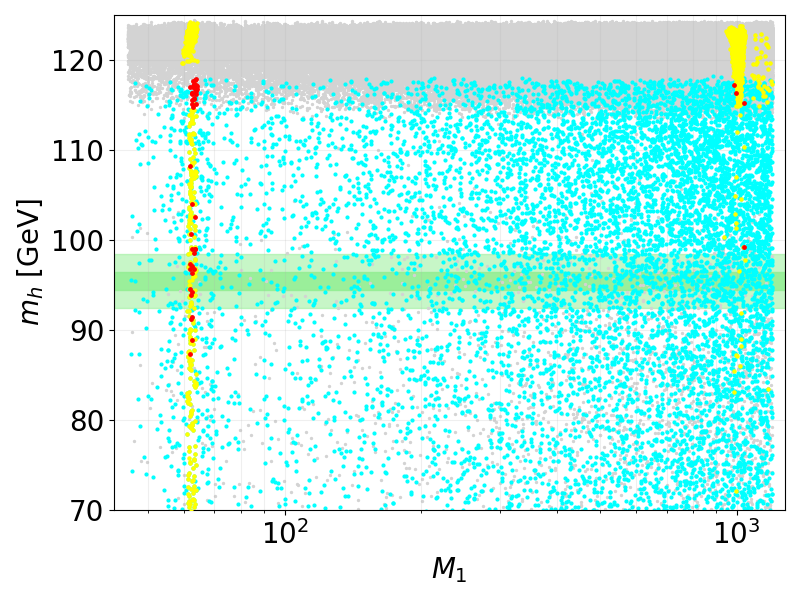}
\includegraphics[width=0.49\textwidth]{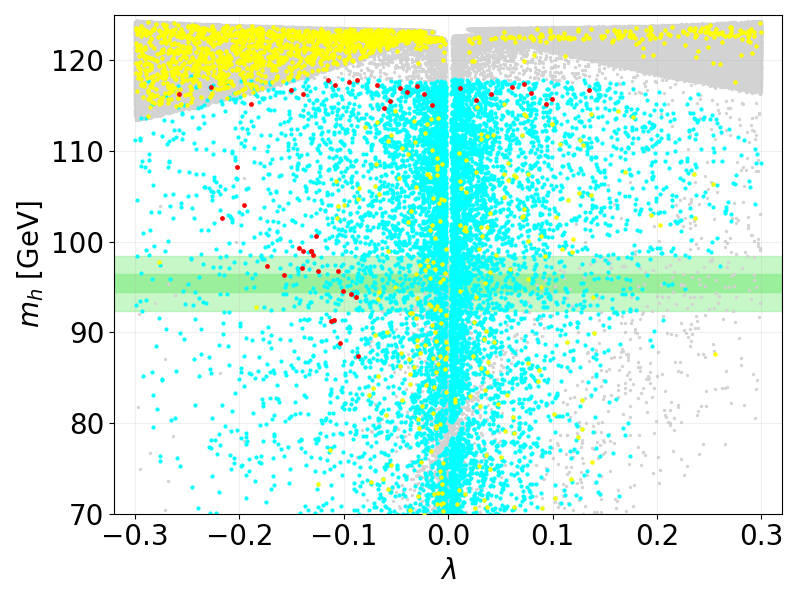}
\caption{The scanning results using the parameter ranges in Table~\ref{tab:TMSSM_parameters} are shown. The left panel displays the relationship between the light scalar Higgs boson mass \( m_h \) and \( M_1 \), while the right panel presents the correlation between \( m_h \) and $\lambda$. The color scheme follows the description in Eq~\ref{eqn:color} and the surrounding context.The green band indicates that $m_h=95.4\pm 1~\mathrm {GeV}   $ , the light green area represents $m_h=95.4\pm 3~\mathrm {GeV}   $.}
\label{fig:NU1}
\end{figure*}

\begin{figure}
    \centering
    \includegraphics[width=1.03\linewidth]{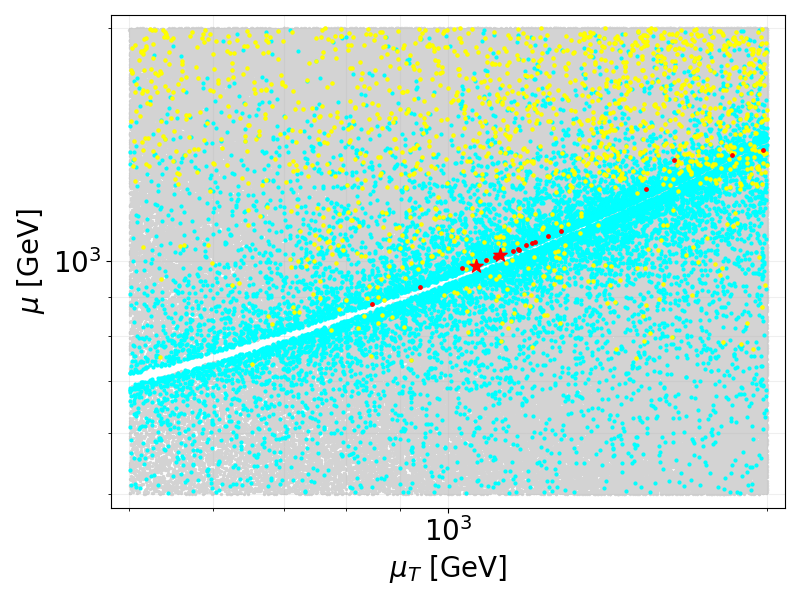}
    \caption{The results of the global scan are presented in the $\mu-\mu_T$ plane. The color coding of the parameter points follows Equation~\ref{eqn:color} and the text below it. The red star points are based on the red points, with the additional condition of $m_h=95.4\pm 1~\mathrm{GeV}$.}.
    \label{fig:NU2}
\end{figure}

We constructed the model’s Lagrangian using SARAH~v4.15.3~\cite{Staub:2008uz} and calculated the particle spectrum, including Higgs-mass loop corrections and diphoton signal strengths, with SPheno~v4.0.5~\cite{Porod:2003um,Porod:2011nf}. Theoretical consistency and experimental Higgs constraints were enforced via HiggsTools~\cite{Bahl:2022igd}, HiggsBounds-5~\cite{Bechtle:2008jh,Bechtle:2011sb,Bechtle:2012lvg,Bechtle:2013wla,Bechtle:2015pma,Bechtle:2020pkv,Bahl:2021yhk}, and HiggsSignals-2~\cite{Bechtle:2013xfa,Stal:2013hwa,Bechtle:2014ewa,Bechtle:2020uwn}, while the DM relic density and DM-nucleon cross-sections were computed with MicrOmegas~v6.1.15~\cite{Belanger:2006is,Belanger:2008sj,Belanger:2010pz,Belanger:2013oya,Belanger:2014vza,Belanger:2018ccd}. Collider limits were validated through SModelS~v3~\cite{smodels:v3}, and the entire workflow was automated using BSMArt~v1.5~\cite{Goodsell:2023iac,Faraggi:2023jzm,Fuks:2024qdt,Agin:2023yoq,Agin:2024yfs,Benakli:2022gjn,Goodsell:2022beo,Goodsell:2021iwc,Goodsell:2020rfu,Domenech:2020yjf}. Fixed parameters are specified and Bayesian parameter estimation is performed using the MultiNest~v3.12~\cite{Feroz:2007kg,Feroz:2008xx,Feroz:2013hea} scanning the free parameter ranges given in Table~\ref{tab:TMSSM_parameters}.

\begin{table}[hbt]
\centering
\renewcommand{\arraystretch}{1.2}
\setlength{\tabcolsep}{0.8em}
\begin{tabular}{l r @{\,}l || l r @{\,}l} 
\hline
\hline
\multicolumn{6}{c}{\textbf{Fixed Parameters}} \\ 
\hline
$M_S$/TeV       & 5   &          & $\tan\beta$ & 3.95 &          \\
$A_0$/TeV       & 12  &          & $M_2$/TeV       & 1.5  &          \\
$M_3$/TeV       & 3   &          & $A_\lambda$/TeV       & 2    &          \\
$B_{\mu_T}$/TeV$^2$ & 5   &           & $B_\mu$/TeV$^2$ & 1    &           \\ 
\hline
\multicolumn{6}{c}{\textbf{Scanned Parameter Ranges}} \\ 
\hline
$M_1$/GeV       & (45, & 1200)      & $\lambda$ & ($-0.3$, & 0.3)    \\
$\mu_T$/GeV     & (500, & 2000)      & $\mu$/GeV     & (500,   & 2000) \\
\hline
\hline
\end{tabular}
\caption{\label{tab:TMSSM_parameters} 
Parameter space of the TMSSM: Fixed values (top) and scanned ranges (bottom). }
\end{table}

Before presenting our results, we define the conditions for the final selected parameter points. To address Higgs mass uncertainties from loop corrections, we set the lightest Higgs boson mass range as $94.4 \ \mathrm{GeV} \leq m_{h} \leq 96.4 \ \mathrm{GeV}$, with the Standard Model Higgs boson mass at $123 \ \mathrm{GeV} \leq m_{H} \leq 127 \ \mathrm{GeV}$. For Higgs data fitting, we incorporate direct search results for additional Higgs bosons from LEP, Tevatron, and LHC experiments. Selected parameter points must satisfy $\mathrm{result} = 1$ in HiggsBounds with the HiggsSignal $p$-value obtained via $\chi^2$ analysis exceeding 0.05 to ensure statistical significance and experimental consistency.

To account for the systematic uncertainties in DM calculations, we adopt an expanded relic density range of $\Omega h^{2} \in \left [ 0.096,0.144 \right ] $, corresponding to $\pm 20\%$ variations around the Planck-2018 observed cold DM central value $\Omega h^{2} =0.120$~\cite{Planck:2018vyg}. For direct and indirect detection constraints, we require a DM signal exclusion statistical reliability of $p \ge   0.05$ at the 95\% confidence level, while collider limits are enforced via the SModelS exclusion ratio criterion $r < 1$~\cite{smodels:v3}.

We additionally require a lower bound $m_{H^\pm} > 110\;\mathrm{GeV}$ on the charged Higgs mass from stau ($\tilde{\tau}$) searches, following the analysis of Ref.~\cite{Ashanujjaman:2024lnr}. The similar final-state signatures between $\tilde{\tau}^\pm \to \tau^\pm \tilde{\chi}^0_1$ and $H^\pm \to \tau\nu$ decays (for massless neutrinos) allow reinterpretation of ATLAS stau search results. The current exclusion limits rule out charged triplet scalars below $110\;\mathrm{GeV}$.

The color scheme employed in the scatter plot adheres to the following consistent convention:

\begin{itemize} 
    \item Red points:
\begin{equation}
    \begin{aligned}
        & m_H = 125 \pm 2~\mathrm{GeV}, \quad \mathrm{HBresult} = 1,  \\
        & \mathrm{HSpval} \geq 0.05, \quad \Omega h^2 = 0.120 \pm 0.024,  \\
        & \mathrm{DMpval} \geq 0.05, \quad \mathrm{SModelSr} < 1,  \\
        & \mu_{\gamma\gamma} = 0.24_{-0.08}^{+0.09}, \quad{m_{H}^{\pm } \ge 110 ~\mathrm {GeV}   }
    \end{aligned}
    \label{eqn:color}
\end{equation}

    \item Cyan points (Higgs constraints): SM Higgs mass lies within the experimentally allowed range and satisfies all constraints from HB, HS, and SModelS.
    
    \item Yellow points (DM constraints): $\Omega {h}^{2}$  lies within a $\pm20\%$ range, DM p-value satisfies statistical requirements and $\mathrm{SModelSr} <  1$.

    \item Gray points: Only $\mathrm{SModelSr} < 1$.
\end{itemize}

The left panel of Fig.~\ref{fig:NU1} shows that the cyan data points are nearly independent of $M_1$. This is because $M_1$ only affects the bino mass, which plays no role in the Higgs sector. In contrast, the yellow data points display a specific distribution, where the first cluster lies in the Higgs resonance region at $M_1 \approx m_H/2$. The second cluster appears near the TeV energy scale. This pattern matches our earlier findings reported in Fig.~\ref{fig:DM1}. In the low-energy Higgs resonance region, the bino mass $M_1$ is much smaller than the higgsino mass $\mu$ and the triplino mass $2\mu_T$. This enables the region to yield a pure bino dark matter candidate consistent with observational constraints. At the TeV scale, the near degeneracy of $M_1 \sim \mu \sim \mu_T$  facilitates bino-higgsino or bino-triplino coannihilation via gauge-mediated interactions. This configuration boosts the efficiency of annihilation, yielding a correct DM relic density. Diphoton excess further reduces the yellow points to fewer red points. Then we can find the overall constraint still has survival points touching the correct 95 GeV Higgs band.

In the right panel of Fig.\ref{fig:NU1}, the quartic coupling $\lambda$ exhibits a $Z_2$ symmetry $\lambda \rightarrow -\lambda$ because it enters the Higgs and DM observables only through $\lambda^2$. Most points satisfying the Higgs constraints have $m_h > 110,\mathrm{GeV}$. The yellow points are predominantly in the negative $\lambda$ region due to the collective influence of parameters such as $M_1$, $\mu$, and $\mu_T$, rather than a direct effect of $\lambda$ itself. As $\lambda$ decreases, the density of points with $m_h < 110,\mathrm{GeV}$ increases, reflecting the interplay of $\lambda$, $\mu$, and $\mu_T$ in the expression for $m_h$ (Eq.\ref{eqn:expression}). Notably, some red points lie within the theoretically predicted green band to match the $95$ GeV Higgs requirement.

Finally, in Fig.\ref{fig:NU2}, the cyan points satisfying the Higgs constraints are diffusely distributed, showing no significant correlation with $\mu$ or $\mu_T$. This indicates that the Higgs constraints impose relatively weak limits on these mass parameters. The yellow points, which satisfy the dark matter constraints, depend on the higgsino mass parameter $\mu$, with their density increasing as $\mu$ increases. This occurs because the bino can become the LSP and achieve the correct relic density only for larger $\mu$. The red points, indicating constraints on the light scalar Higgs boson mass $m_h$ and signal strength, show a linear dependence on $\mu$ and $\mu_T$. This narrow band results from the cancellation between these mass parameters, consistent with Fig.\ref{fig:Higgs95.4} and Eq.\ref{eqn:expression}. The jagged white regions represent parameter space where the triplino is the next-to-lightest stable particle, consistent with Fig.\ref{fig:LSP}.

\section{Conclusions}
\label{s5}

This study systematically explores the TMSSM to explain the $95$ GeV Higgs boson diphoton excess observed by CMS and ATLAS and its implications for DM physics. By incorporating an $SU(2)_L$ triplet superfield, TMSSM reduces fine-tuning issues of the MSSM and naturally supports Higgs boson masses at $95$ GeV and $125$ GeV. Our calculations show that optimizing triplet-doublet mixing and parameters $\lambda$ and $\mu_T$ enables TMSSM to reproduce the experimental diphoton signal strength, $\mu_{\gamma\gamma}^{\text{CMS+ATLAS}} = 0.24_{-0.08}^{+0.09}$. In the DM sector, the triplet and its supersymmetric partner, the triplino, enhance neutralino annihilation through Higgs resonance and co-annihilation, achieving consistency with the Planck-2018 cold DM relic density, $\Omega h^2 = 0.120 \pm 0.024$, while meeting direct and indirect detection constraints.

Using tools like SARAH, SPheno, MicrOmegas, and BSMArts, we performed a comprehensive scan of the TMSSM parameter space, identifying regions that simultaneously satisfy Higgs masses, $m_h \approx 95.4 \pm 1$ GeV and $m_H \approx 125 \pm 2$ GeV, diphoton signal strength, and DM relic density requirements. These findings confirm that TMSSM robustly explains the 95 GeV diphoton excess and provides a consistent framework for DM physics. Future collider experiments, such as the High-Luminosity LHC or a next-generation circular collider, will precisely probe triplino properties and the 95 GeV Higgs signal, offering critical tests of TMSSM and deeper insights into supersymmetric physics.

\begin{acknowledgments}
We thank Mark Goodsell and Andreas Crivellin for the helpful discussions. This work is supported by the National Natural Science Foundation of China (NNSFC) under grant Nos. 12275232, No. 12335005.
\end{acknowledgments}
 
\vspace{-.3cm}
\bibliography{main}

\begin{thebibliography}{121}%
\makeatletter
\providecommand \@ifxundefined [1]{%
 \@ifx{#1\undefined}
}%
\providecommand \@ifnum [1]{%
 \ifnum #1\expandafter \@firstoftwo
 \else \expandafter \@secondoftwo
 \fi
}%
\providecommand \@ifx [1]{%
 \ifx #1\expandafter \@firstoftwo
 \else \expandafter \@secondoftwo
 \fi
}%
\providecommand \natexlab [1]{#1}%
\providecommand \enquote  [1]{``#1''}%
\providecommand \bibnamefont  [1]{#1}%
\providecommand \bibfnamefont [1]{#1}%
\providecommand \citenamefont [1]{#1}%
\providecommand \href@noop [0]{\@secondoftwo}%
\providecommand \href [0]{\begingroup \@sanitize@url \@href}%
\providecommand \@href[1]{\@@startlink{#1}\@@href}%
\providecommand \@@href[1]{\endgroup#1\@@endlink}%
\providecommand \@sanitize@url [0]{\catcode `\\12\catcode `\$12\catcode `\&12\catcode `\#12\catcode `\^12\catcode `\_12\catcode `\%12\relax}%
\providecommand \@@startlink[1]{}%
\providecommand \@@endlink[0]{}%
\providecommand \url  [0]{\begingroup\@sanitize@url \@url }%
\providecommand \@url [1]{\endgroup\@href {#1}{\urlprefix }}%
\providecommand \urlprefix  [0]{URL }%
\providecommand \Eprint [0]{\href }%
\providecommand \doibase [0]{http://dx.doi.org/}%
\providecommand \selectlanguage [0]{\@gobble}%
\providecommand \bibinfo  [0]{\@secondoftwo}%
\providecommand \bibfield  [0]{\@secondoftwo}%
\providecommand \translation [1]{[#1]}%
\providecommand \BibitemOpen [0]{}%
\providecommand \bibitemStop [0]{}%
\providecommand \bibitemNoStop [0]{.\EOS\space}%
\providecommand \EOS [0]{\spacefactor3000\relax}%
\providecommand \BibitemShut  [1]{\csname bibitem#1\endcsname}%
\let\auto@bib@innerbib\@empty
\bibitem [{\citenamefont {Randall}\ and\ \citenamefont {Sundrum}(1999)}]{Randall:1999ee}%
  \BibitemOpen
  \bibfield  {author} {\bibinfo {author} {\bibfnamefont {L.}~\bibnamefont {Randall}}\ and\ \bibinfo {author} {\bibfnamefont {R.}~\bibnamefont {Sundrum}},\ }\href {\doibase 10.1103/PhysRevLett.83.3370} {\bibfield  {journal} {\bibinfo  {journal} {Phys. Rev. Lett.}\ }\textbf {\bibinfo {volume} {83}},\ \bibinfo {pages} {3370} (\bibinfo {year} {1999})},\ \Eprint {http://arxiv.org/abs/hep-ph/9905221} {arXiv:hep-ph/9905221} \BibitemShut {NoStop}%
\bibitem [{\citenamefont {Arkani-Hamed}\ \emph {et~al.}(1998)\citenamefont {Arkani-Hamed}, \citenamefont {Dimopoulos},\ and\ \citenamefont {Dvali}}]{Arkani-Hamed:1998jmv}%
  \BibitemOpen
  \bibfield  {author} {\bibinfo {author} {\bibfnamefont {N.}~\bibnamefont {Arkani-Hamed}}, \bibinfo {author} {\bibfnamefont {S.}~\bibnamefont {Dimopoulos}}, \ and\ \bibinfo {author} {\bibfnamefont {G.~R.}\ \bibnamefont {Dvali}},\ }\href {\doibase 10.1016/S0370-2693(98)00466-3} {\bibfield  {journal} {\bibinfo  {journal} {Phys. Lett. B}\ }\textbf {\bibinfo {volume} {429}},\ \bibinfo {pages} {263} (\bibinfo {year} {1998})},\ \Eprint {http://arxiv.org/abs/hep-ph/9803315} {arXiv:hep-ph/9803315} \BibitemShut {NoStop}%
\bibitem [{\citenamefont {Haber}\ and\ \citenamefont {Kane}(1985)}]{Haber:1984rc}%
  \BibitemOpen
  \bibfield  {author} {\bibinfo {author} {\bibfnamefont {H.~E.}\ \bibnamefont {Haber}}\ and\ \bibinfo {author} {\bibfnamefont {G.~L.}\ \bibnamefont {Kane}},\ }\href {\doibase 10.1016/0370-1573(85)90051-1} {\bibfield  {journal} {\bibinfo  {journal} {Phys. Rept.}\ }\textbf {\bibinfo {volume} {117}},\ \bibinfo {pages} {75} (\bibinfo {year} {1985})}\BibitemShut {NoStop}%
\bibitem [{\citenamefont {Martin}(1998)}]{Martin:1997ns}%
  \BibitemOpen
  \bibfield  {author} {\bibinfo {author} {\bibfnamefont {S.~P.}\ \bibnamefont {Martin}},\ }\href {\doibase 10.1142/9789812839657_0001} {\bibfield  {journal} {\bibinfo  {journal} {Adv. Ser. Direct. High Energy Phys.}\ }\textbf {\bibinfo {volume} {18}},\ \bibinfo {pages} {1} (\bibinfo {year} {1998})},\ \Eprint {http://arxiv.org/abs/hep-ph/9709356} {arXiv:hep-ph/9709356} \BibitemShut {NoStop}%
\bibitem [{\citenamefont {Jungman}\ \emph {et~al.}(1996)\citenamefont {Jungman}, \citenamefont {Kamionkowski},\ and\ \citenamefont {Griest}}]{Jungman:1995df}%
  \BibitemOpen
  \bibfield  {author} {\bibinfo {author} {\bibfnamefont {G.}~\bibnamefont {Jungman}}, \bibinfo {author} {\bibfnamefont {M.}~\bibnamefont {Kamionkowski}}, \ and\ \bibinfo {author} {\bibfnamefont {K.}~\bibnamefont {Griest}},\ }\href {\doibase 10.1016/0370-1573(95)00058-5} {\bibfield  {journal} {\bibinfo  {journal} {Phys. Rept.}\ }\textbf {\bibinfo {volume} {267}},\ \bibinfo {pages} {195} (\bibinfo {year} {1996})},\ \Eprint {http://arxiv.org/abs/hep-ph/9506380} {arXiv:hep-ph/9506380} \BibitemShut {NoStop}%
\bibitem [{\citenamefont {Duan}\ \emph {et~al.}(2018{\natexlab{a}})\citenamefont {Duan}, \citenamefont {Wang}, \citenamefont {Wu}, \citenamefont {Yang},\ and\ \citenamefont {Zhao}}]{Duan:2017ucw}%
  \BibitemOpen
  \bibfield  {author} {\bibinfo {author} {\bibfnamefont {G.~H.}\ \bibnamefont {Duan}}, \bibinfo {author} {\bibfnamefont {W.}~\bibnamefont {Wang}}, \bibinfo {author} {\bibfnamefont {L.}~\bibnamefont {Wu}}, \bibinfo {author} {\bibfnamefont {J.~M.}\ \bibnamefont {Yang}}, \ and\ \bibinfo {author} {\bibfnamefont {J.}~\bibnamefont {Zhao}},\ }\href {\doibase 10.1016/j.physletb.2018.01.030} {\bibfield  {journal} {\bibinfo  {journal} {Phys. Lett. B}\ }\textbf {\bibinfo {volume} {778}},\ \bibinfo {pages} {296} (\bibinfo {year} {2018}{\natexlab{a}})},\ \Eprint {http://arxiv.org/abs/1711.03893} {arXiv:1711.03893 [hep-ph]} \BibitemShut {NoStop}%
\bibitem [{\citenamefont {Kitano}\ and\ \citenamefont {Nomura}(2006)}]{Kitano:2006gv}%
  \BibitemOpen
  \bibfield  {author} {\bibinfo {author} {\bibfnamefont {R.}~\bibnamefont {Kitano}}\ and\ \bibinfo {author} {\bibfnamefont {Y.}~\bibnamefont {Nomura}},\ }\href {\doibase 10.1103/PhysRevD.73.095004} {\bibfield  {journal} {\bibinfo  {journal} {Phys. Rev. D}\ }\textbf {\bibinfo {volume} {73}},\ \bibinfo {pages} {095004} (\bibinfo {year} {2006})},\ \Eprint {http://arxiv.org/abs/hep-ph/0602096} {arXiv:hep-ph/0602096} \BibitemShut {NoStop}%
\bibitem [{\citenamefont {Papucci}\ \emph {et~al.}(2012)\citenamefont {Papucci}, \citenamefont {Ruderman},\ and\ \citenamefont {Weiler}}]{Papucci:2011wy}%
  \BibitemOpen
  \bibfield  {author} {\bibinfo {author} {\bibfnamefont {M.}~\bibnamefont {Papucci}}, \bibinfo {author} {\bibfnamefont {J.~T.}\ \bibnamefont {Ruderman}}, \ and\ \bibinfo {author} {\bibfnamefont {A.}~\bibnamefont {Weiler}},\ }\href {\doibase 10.1007/JHEP09(2012)035} {\bibfield  {journal} {\bibinfo  {journal} {JHEP}\ }\textbf {\bibinfo {volume} {09}},\ \bibinfo {pages} {035} (\bibinfo {year} {2012})},\ \Eprint {http://arxiv.org/abs/1110.6926} {arXiv:1110.6926 [hep-ph]} \BibitemShut {NoStop}%
\bibitem [{\citenamefont {Strumia}(2011)}]{Strumia:2011dv}%
  \BibitemOpen
  \bibfield  {author} {\bibinfo {author} {\bibfnamefont {A.}~\bibnamefont {Strumia}},\ }\href {\doibase 10.1007/JHEP04(2011)073} {\bibfield  {journal} {\bibinfo  {journal} {JHEP}\ }\textbf {\bibinfo {volume} {04}},\ \bibinfo {pages} {073} (\bibinfo {year} {2011})},\ \Eprint {http://arxiv.org/abs/1101.2195} {arXiv:1101.2195 [hep-ph]} \BibitemShut {NoStop}%
\bibitem [{\citenamefont {Craig}\ \emph {et~al.}(2015)\citenamefont {Craig}, \citenamefont {Katz}, \citenamefont {Strassler},\ and\ \citenamefont {Sundrum}}]{Craig:2015pha}%
  \BibitemOpen
  \bibfield  {author} {\bibinfo {author} {\bibfnamefont {N.}~\bibnamefont {Craig}}, \bibinfo {author} {\bibfnamefont {A.}~\bibnamefont {Katz}}, \bibinfo {author} {\bibfnamefont {M.}~\bibnamefont {Strassler}}, \ and\ \bibinfo {author} {\bibfnamefont {R.}~\bibnamefont {Sundrum}},\ }\href {\doibase 10.1007/JHEP07(2015)105} {\bibfield  {journal} {\bibinfo  {journal} {JHEP}\ }\textbf {\bibinfo {volume} {07}},\ \bibinfo {pages} {105} (\bibinfo {year} {2015})},\ \Eprint {http://arxiv.org/abs/1501.05310} {arXiv:1501.05310 [hep-ph]} \BibitemShut {NoStop}%
\bibitem [{\citenamefont {Hayrapetyan}\ \emph {et~al.}(2024)\citenamefont {Hayrapetyan} \emph {et~al.}}]{CMS:2023mny}%
  \BibitemOpen
  \bibfield  {author} {\bibinfo {author} {\bibfnamefont {A.}~\bibnamefont {Hayrapetyan}} \emph {et~al.} (\bibinfo {collaboration} {CMS}),\ }\href {\doibase 10.1103/PhysRevD.109.072007} {\bibfield  {journal} {\bibinfo  {journal} {Phys. Rev. D}\ }\textbf {\bibinfo {volume} {109}},\ \bibinfo {pages} {072007} (\bibinfo {year} {2024})},\ \Eprint {http://arxiv.org/abs/2309.16823} {arXiv:2309.16823 [hep-ex]} \BibitemShut {NoStop}%
\bibitem [{\citenamefont {Aad}\ \emph {et~al.}(2025)\citenamefont {Aad} \emph {et~al.}}]{ATLAS:2024lda}%
  \BibitemOpen
  \bibfield  {author} {\bibinfo {author} {\bibfnamefont {G.}~\bibnamefont {Aad}} \emph {et~al.} (\bibinfo {collaboration} {ATLAS}),\ }\href {\doibase 10.1016/j.physrep.2024.09.010} {\bibfield  {journal} {\bibinfo  {journal} {Phys. Rept.}\ }\textbf {\bibinfo {volume} {1116}},\ \bibinfo {pages} {261} (\bibinfo {year} {2025})},\ \Eprint {http://arxiv.org/abs/2403.02455} {arXiv:2403.02455 [hep-ex]} \BibitemShut {NoStop}%
\bibitem [{\citenamefont {Espinosa}\ and\ \citenamefont {Quiros}(1992)}]{Espinosa:1991wt}%
  \BibitemOpen
  \bibfield  {author} {\bibinfo {author} {\bibfnamefont {J.~R.}\ \bibnamefont {Espinosa}}\ and\ \bibinfo {author} {\bibfnamefont {M.}~\bibnamefont {Quiros}},\ }\href {\doibase 10.1016/0550-3213(92)90464-M} {\bibfield  {journal} {\bibinfo  {journal} {Nucl. Phys. B}\ }\textbf {\bibinfo {volume} {384}},\ \bibinfo {pages} {113} (\bibinfo {year} {1992})}\BibitemShut {NoStop}%
\bibitem [{\citenamefont {Di~Chiara}\ and\ \citenamefont {Hsieh}(2008)}]{DiChiara:2008rg}%
  \BibitemOpen
  \bibfield  {author} {\bibinfo {author} {\bibfnamefont {S.}~\bibnamefont {Di~Chiara}}\ and\ \bibinfo {author} {\bibfnamefont {K.}~\bibnamefont {Hsieh}},\ }\href {\doibase 10.1103/PhysRevD.78.055016} {\bibfield  {journal} {\bibinfo  {journal} {Phys. Rev. D}\ }\textbf {\bibinfo {volume} {78}},\ \bibinfo {pages} {055016} (\bibinfo {year} {2008})},\ \Eprint {http://arxiv.org/abs/0805.2623} {arXiv:0805.2623 [hep-ph]} \BibitemShut {NoStop}%
\bibitem [{\citenamefont {Basak}\ and\ \citenamefont {Mohanty}(2012)}]{Basak:2012bd}%
  \BibitemOpen
  \bibfield  {author} {\bibinfo {author} {\bibfnamefont {T.}~\bibnamefont {Basak}}\ and\ \bibinfo {author} {\bibfnamefont {S.}~\bibnamefont {Mohanty}},\ }\href {\doibase 10.1103/PhysRevD.86.075031} {\bibfield  {journal} {\bibinfo  {journal} {Phys. Rev. D}\ }\textbf {\bibinfo {volume} {86}},\ \bibinfo {pages} {075031} (\bibinfo {year} {2012})},\ \Eprint {http://arxiv.org/abs/1204.6592} {arXiv:1204.6592 [hep-ph]} \BibitemShut {NoStop}%
\bibitem [{\citenamefont {Cort}\ \emph {et~al.}(2013)\citenamefont {Cort}, \citenamefont {Garcia},\ and\ \citenamefont {Quiros}}]{Cort:2013foa}%
  \BibitemOpen
  \bibfield  {author} {\bibinfo {author} {\bibfnamefont {L.}~\bibnamefont {Cort}}, \bibinfo {author} {\bibfnamefont {M.}~\bibnamefont {Garcia}}, \ and\ \bibinfo {author} {\bibfnamefont {M.}~\bibnamefont {Quiros}},\ }\href {\doibase 10.1103/PhysRevD.88.075010} {\bibfield  {journal} {\bibinfo  {journal} {Phys. Rev. D}\ }\textbf {\bibinfo {volume} {88}},\ \bibinfo {pages} {075010} (\bibinfo {year} {2013})},\ \Eprint {http://arxiv.org/abs/1308.4025} {arXiv:1308.4025 [hep-ph]} \BibitemShut {NoStop}%
\bibitem [{\citenamefont {Bandyopadhyay}\ \emph {et~al.}(2013)\citenamefont {Bandyopadhyay}, \citenamefont {Huitu},\ and\ \citenamefont {Sabanci}}]{Bandyopadhyay:2013lca}%
  \BibitemOpen
  \bibfield  {author} {\bibinfo {author} {\bibfnamefont {P.}~\bibnamefont {Bandyopadhyay}}, \bibinfo {author} {\bibfnamefont {K.}~\bibnamefont {Huitu}}, \ and\ \bibinfo {author} {\bibfnamefont {A.}~\bibnamefont {Sabanci}},\ }\href {\doibase 10.1007/JHEP10(2013)091} {\bibfield  {journal} {\bibinfo  {journal} {JHEP}\ }\textbf {\bibinfo {volume} {10}},\ \bibinfo {pages} {091} (\bibinfo {year} {2013})},\ \Eprint {http://arxiv.org/abs/1306.4530} {arXiv:1306.4530 [hep-ph]} \BibitemShut {NoStop}%
\bibitem [{\citenamefont {Bandyopadhyay}\ \emph {et~al.}(2014)\citenamefont {Bandyopadhyay}, \citenamefont {Di~Chiara}, \citenamefont {Huitu},\ and\ \citenamefont {Ke\c{c}eli}}]{Bandyopadhyay:2014tha}%
  \BibitemOpen
  \bibfield  {author} {\bibinfo {author} {\bibfnamefont {P.}~\bibnamefont {Bandyopadhyay}}, \bibinfo {author} {\bibfnamefont {S.}~\bibnamefont {Di~Chiara}}, \bibinfo {author} {\bibfnamefont {K.}~\bibnamefont {Huitu}}, \ and\ \bibinfo {author} {\bibfnamefont {A.~S.}\ \bibnamefont {Ke\c{c}eli}},\ }\href {\doibase 10.1007/JHEP11(2014)062} {\bibfield  {journal} {\bibinfo  {journal} {JHEP}\ }\textbf {\bibinfo {volume} {11}},\ \bibinfo {pages} {062} (\bibinfo {year} {2014})},\ \Eprint {http://arxiv.org/abs/1407.4836} {arXiv:1407.4836 [hep-ph]} \BibitemShut {NoStop}%
\bibitem [{\citenamefont {Garcia-Pepin}\ \emph {et~al.}(2015)\citenamefont {Garcia-Pepin}, \citenamefont {Gori}, \citenamefont {Quiros}, \citenamefont {Vega}, \citenamefont {Vega-Morales},\ and\ \citenamefont {Yu}}]{Garcia-Pepin:2014yfa}%
  \BibitemOpen
  \bibfield  {author} {\bibinfo {author} {\bibfnamefont {M.}~\bibnamefont {Garcia-Pepin}}, \bibinfo {author} {\bibfnamefont {S.}~\bibnamefont {Gori}}, \bibinfo {author} {\bibfnamefont {M.}~\bibnamefont {Quiros}}, \bibinfo {author} {\bibfnamefont {R.}~\bibnamefont {Vega}}, \bibinfo {author} {\bibfnamefont {R.}~\bibnamefont {Vega-Morales}}, \ and\ \bibinfo {author} {\bibfnamefont {T.-T.}\ \bibnamefont {Yu}},\ }\href {\doibase 10.1103/PhysRevD.91.015016} {\bibfield  {journal} {\bibinfo  {journal} {Phys. Rev. D}\ }\textbf {\bibinfo {volume} {91}},\ \bibinfo {pages} {015016} (\bibinfo {year} {2015})},\ \Eprint {http://arxiv.org/abs/1409.5737} {arXiv:1409.5737 [hep-ph]} \BibitemShut {NoStop}%
\bibitem [{\citenamefont {Bandyopadhyay}\ \emph {et~al.}(2015{\natexlab{a}})\citenamefont {Bandyopadhyay}, \citenamefont {Coriano},\ and\ \citenamefont {Costantini}}]{Bandyopadhyay:2015oga}%
  \BibitemOpen
  \bibfield  {author} {\bibinfo {author} {\bibfnamefont {P.}~\bibnamefont {Bandyopadhyay}}, \bibinfo {author} {\bibfnamefont {C.}~\bibnamefont {Coriano}}, \ and\ \bibinfo {author} {\bibfnamefont {A.}~\bibnamefont {Costantini}},\ }\href {\doibase 10.1007/JHEP09(2015)045} {\bibfield  {journal} {\bibinfo  {journal} {JHEP}\ }\textbf {\bibinfo {volume} {09}},\ \bibinfo {pages} {045} (\bibinfo {year} {2015}{\natexlab{a}})},\ \Eprint {http://arxiv.org/abs/1506.03634} {arXiv:1506.03634 [hep-ph]} \BibitemShut {NoStop}%
\bibitem [{\citenamefont {Bandyopadhyay}\ \emph {et~al.}(2016)\citenamefont {Bandyopadhyay}, \citenamefont {Corian\`o},\ and\ \citenamefont {Costantini}}]{Bandyopadhyay:2015ifm}%
  \BibitemOpen
  \bibfield  {author} {\bibinfo {author} {\bibfnamefont {P.}~\bibnamefont {Bandyopadhyay}}, \bibinfo {author} {\bibfnamefont {C.}~\bibnamefont {Corian\`o}}, \ and\ \bibinfo {author} {\bibfnamefont {A.}~\bibnamefont {Costantini}},\ }\href {\doibase 10.1103/PhysRevD.94.055030} {\bibfield  {journal} {\bibinfo  {journal} {Phys. Rev. D}\ }\textbf {\bibinfo {volume} {94}},\ \bibinfo {pages} {055030} (\bibinfo {year} {2016})},\ \Eprint {http://arxiv.org/abs/1512.08651} {arXiv:1512.08651 [hep-ph]} \BibitemShut {NoStop}%
\bibitem [{\citenamefont {Bandyopadhyay}\ \emph {et~al.}(2015{\natexlab{b}})\citenamefont {Bandyopadhyay}, \citenamefont {Coriano},\ and\ \citenamefont {Costantini}}]{Bandyopadhyay:2015tva}%
  \BibitemOpen
  \bibfield  {author} {\bibinfo {author} {\bibfnamefont {P.}~\bibnamefont {Bandyopadhyay}}, \bibinfo {author} {\bibfnamefont {C.}~\bibnamefont {Coriano}}, \ and\ \bibinfo {author} {\bibfnamefont {A.}~\bibnamefont {Costantini}},\ }\href {\doibase 10.1007/JHEP12(2015)127} {\bibfield  {journal} {\bibinfo  {journal} {JHEP}\ }\textbf {\bibinfo {volume} {12}},\ \bibinfo {pages} {127} (\bibinfo {year} {2015}{\natexlab{b}})},\ \Eprint {http://arxiv.org/abs/1510.06309} {arXiv:1510.06309 [hep-ph]} \BibitemShut {NoStop}%
\bibitem [{\citenamefont {Diaz-Cruz}\ \emph {et~al.}(2008)\citenamefont {Diaz-Cruz}, \citenamefont {Hernandez-Sanchez}, \citenamefont {Moretti},\ and\ \citenamefont {Rosado}}]{Diaz-Cruz:2007rcq}%
  \BibitemOpen
  \bibfield  {author} {\bibinfo {author} {\bibfnamefont {J.~L.}\ \bibnamefont {Diaz-Cruz}}, \bibinfo {author} {\bibfnamefont {J.}~\bibnamefont {Hernandez-Sanchez}}, \bibinfo {author} {\bibfnamefont {S.}~\bibnamefont {Moretti}}, \ and\ \bibinfo {author} {\bibfnamefont {A.}~\bibnamefont {Rosado}},\ }\href {\doibase 10.1103/PhysRevD.77.035007} {\bibfield  {journal} {\bibinfo  {journal} {Phys. Rev. D}\ }\textbf {\bibinfo {volume} {77}},\ \bibinfo {pages} {035007} (\bibinfo {year} {2008})},\ \Eprint {http://arxiv.org/abs/0710.4169} {arXiv:0710.4169 [hep-ph]} \BibitemShut {NoStop}%
\bibitem [{\citenamefont {Arina}\ \emph {et~al.}(2014)\citenamefont {Arina}, \citenamefont {Martin-Lozano},\ and\ \citenamefont {Nardini}}]{Arina:2014xya}%
  \BibitemOpen
  \bibfield  {author} {\bibinfo {author} {\bibfnamefont {C.}~\bibnamefont {Arina}}, \bibinfo {author} {\bibfnamefont {V.}~\bibnamefont {Martin-Lozano}}, \ and\ \bibinfo {author} {\bibfnamefont {G.}~\bibnamefont {Nardini}},\ }\href {\doibase 10.1007/JHEP08(2014)015} {\bibfield  {journal} {\bibinfo  {journal} {JHEP}\ }\textbf {\bibinfo {volume} {08}},\ \bibinfo {pages} {015} (\bibinfo {year} {2014})},\ \Eprint {http://arxiv.org/abs/1403.6434} {arXiv:1403.6434 [hep-ph]} \BibitemShut {NoStop}%
\bibitem [{\citenamefont {Barradas-Guevara}\ \emph {et~al.}(2005)\citenamefont {Barradas-Guevara}, \citenamefont {Felix-Beltran},\ and\ \citenamefont {Rosado}}]{Barradas-Guevara:2004all}%
  \BibitemOpen
  \bibfield  {author} {\bibinfo {author} {\bibfnamefont {E.}~\bibnamefont {Barradas-Guevara}}, \bibinfo {author} {\bibfnamefont {O.}~\bibnamefont {Felix-Beltran}}, \ and\ \bibinfo {author} {\bibfnamefont {A.}~\bibnamefont {Rosado}},\ }\href {\doibase 10.1103/PhysRevD.71.073004} {\bibfield  {journal} {\bibinfo  {journal} {Phys. Rev. D}\ }\textbf {\bibinfo {volume} {71}},\ \bibinfo {pages} {073004} (\bibinfo {year} {2005})},\ \Eprint {http://arxiv.org/abs/hep-ph/0408196} {arXiv:hep-ph/0408196} \BibitemShut {NoStop}%
\bibitem [{\citenamefont {Das}\ \emph {et~al.}(2015)\citenamefont {Das}, \citenamefont {Di~Chiara},\ and\ \citenamefont {Roy}}]{Das:2015tva}%
  \BibitemOpen
  \bibfield  {author} {\bibinfo {author} {\bibfnamefont {M.}~\bibnamefont {Das}}, \bibinfo {author} {\bibfnamefont {S.}~\bibnamefont {Di~Chiara}}, \ and\ \bibinfo {author} {\bibfnamefont {S.}~\bibnamefont {Roy}},\ }\href {\doibase 10.1103/PhysRevD.91.055013} {\bibfield  {journal} {\bibinfo  {journal} {Phys. Rev. D}\ }\textbf {\bibinfo {volume} {91}},\ \bibinfo {pages} {055013} (\bibinfo {year} {2015})},\ \Eprint {http://arxiv.org/abs/1501.02667} {arXiv:1501.02667 [hep-ph]} \BibitemShut {NoStop}%
\bibitem [{\citenamefont {Fileviez~Perez}\ and\ \citenamefont {Wise}(2011)}]{FileviezPerez:2011dg}%
  \BibitemOpen
  \bibfield  {author} {\bibinfo {author} {\bibfnamefont {P.}~\bibnamefont {Fileviez~Perez}}\ and\ \bibinfo {author} {\bibfnamefont {M.~B.}\ \bibnamefont {Wise}},\ }\href {\doibase 10.1103/PhysRevD.84.055015} {\bibfield  {journal} {\bibinfo  {journal} {Phys. Rev. D}\ }\textbf {\bibinfo {volume} {84}},\ \bibinfo {pages} {055015} (\bibinfo {year} {2011})},\ \Eprint {http://arxiv.org/abs/1105.3190} {arXiv:1105.3190 [hep-ph]} \BibitemShut {NoStop}%
\bibitem [{\citenamefont {Sirunyan}\ \emph {et~al.}(2019)\citenamefont {Sirunyan} \emph {et~al.}}]{CMS:2018cyk}%
  \BibitemOpen
  \bibfield  {author} {\bibinfo {author} {\bibfnamefont {A.~M.}\ \bibnamefont {Sirunyan}} \emph {et~al.} (\bibinfo {collaboration} {CMS}),\ }\href {\doibase 10.1016/j.physletb.2019.03.064} {\bibfield  {journal} {\bibinfo  {journal} {Phys. Lett. B}\ }\textbf {\bibinfo {volume} {793}},\ \bibinfo {pages} {320} (\bibinfo {year} {2019})},\ \Eprint {http://arxiv.org/abs/1811.08459} {arXiv:1811.08459 [hep-ex]} \BibitemShut {NoStop}%
\bibitem [{\citenamefont {{CMS collaboration}}(2018)}]{CMS:2023yay}%
  \BibitemOpen
  \bibfield  {author} {\bibinfo {author} {\bibnamefont {{CMS collaboration}}} (\bibinfo {collaboration} {CMS}),\ }\href@noop {} {\bibfield  {journal} {\bibinfo  {journal} {{CMS-PAS-HIG-20-002}}\ } (\bibinfo {year} {2018})},\ \bibinfo {note} {{Search for a standard model-like Higgs boson in the mass range between 70 and 110$~\mathrm{GeV}$ in the diphoton final state in proton-proton collisions at $\sqrt{s}=13~\mathrm{TeV}$}}\BibitemShut {NoStop}%
\bibitem [{\citenamefont {{ATLAS collaboration}}(2018)}]{ATLAS:2018xad}%
  \BibitemOpen
  \bibfield  {author} {\bibinfo {author} {\bibnamefont {{ATLAS collaboration}}} (\bibinfo {collaboration} {ATLAS}),\ }\href@noop {} {\bibfield  {journal} {\bibinfo  {journal} {{ATLAS-CONF-2018-025}}\ } (\bibinfo {year} {2018})},\ \bibinfo {note} {{Search for resonances in the 65 to 110 GeV diphoton invariant mass range using 80 fb$^{-1}$ of $pp$ collisions collected at $\sqrt{s}=13$ TeV with the ATLAS detector}}\BibitemShut {NoStop}%
\bibitem [{\citenamefont {{ATLAS collaboration}}(2023)}]{ATLAS:2023jzc}%
  \BibitemOpen
  \bibfield  {author} {\bibinfo {author} {\bibnamefont {{ATLAS collaboration}}} (\bibinfo {collaboration} {ATLAS}),\ }\href@noop {} {\bibfield  {journal} {\bibinfo  {journal} {{ATLAS-CONF-2023-035}}\ } (\bibinfo {year} {2023})},\ \bibinfo {note} {{Search for diphoton resonances in the 66 to 110 GeV mass range using 140 fb$^{-1}$ of 13 TeV $pp$ collisions collected with the ATLAS detector}}\BibitemShut {NoStop}%
\bibitem [{\citenamefont {Biek\"otter}\ \emph {et~al.}(2024)\citenamefont {Biek\"otter}, \citenamefont {Heinemeyer},\ and\ \citenamefont {Weiglein}}]{Biekotter:2023oen}%
  \BibitemOpen
  \bibfield  {author} {\bibinfo {author} {\bibfnamefont {T.}~\bibnamefont {Biek\"otter}}, \bibinfo {author} {\bibfnamefont {S.}~\bibnamefont {Heinemeyer}}, \ and\ \bibinfo {author} {\bibfnamefont {G.}~\bibnamefont {Weiglein}},\ }\href {\doibase 10.1103/PhysRevD.109.035005} {\bibfield  {journal} {\bibinfo  {journal} {Phys. Rev. D}\ }\textbf {\bibinfo {volume} {109}},\ \bibinfo {pages} {035005} (\bibinfo {year} {2024})},\ \Eprint {http://arxiv.org/abs/2306.03889} {arXiv:2306.03889 [hep-ph]} \BibitemShut {NoStop}%
\bibitem [{\citenamefont {Cao}\ \emph {et~al.}(2017)\citenamefont {Cao}, \citenamefont {Guo}, \citenamefont {He}, \citenamefont {Wu},\ and\ \citenamefont {Zhang}}]{Cao:2016uwt}%
  \BibitemOpen
  \bibfield  {author} {\bibinfo {author} {\bibfnamefont {J.}~\bibnamefont {Cao}}, \bibinfo {author} {\bibfnamefont {X.}~\bibnamefont {Guo}}, \bibinfo {author} {\bibfnamefont {Y.}~\bibnamefont {He}}, \bibinfo {author} {\bibfnamefont {P.}~\bibnamefont {Wu}}, \ and\ \bibinfo {author} {\bibfnamefont {Y.}~\bibnamefont {Zhang}},\ }\href {\doibase 10.1103/PhysRevD.95.116001} {\bibfield  {journal} {\bibinfo  {journal} {Phys. Rev. D}\ }\textbf {\bibinfo {volume} {95}},\ \bibinfo {pages} {116001} (\bibinfo {year} {2017})},\ \Eprint {http://arxiv.org/abs/1612.08522} {arXiv:1612.08522 [hep-ph]} \BibitemShut {NoStop}%
\bibitem [{\citenamefont {Crivellin}\ \emph {et~al.}(2018)\citenamefont {Crivellin}, \citenamefont {Heeck},\ and\ \citenamefont {M\"uller}}]{Crivellin:2017upt}%
  \BibitemOpen
  \bibfield  {author} {\bibinfo {author} {\bibfnamefont {A.}~\bibnamefont {Crivellin}}, \bibinfo {author} {\bibfnamefont {J.}~\bibnamefont {Heeck}}, \ and\ \bibinfo {author} {\bibfnamefont {D.}~\bibnamefont {M\"uller}},\ }\href {\doibase 10.1103/PhysRevD.97.035008} {\bibfield  {journal} {\bibinfo  {journal} {Phys. Rev. D}\ }\textbf {\bibinfo {volume} {97}},\ \bibinfo {pages} {035008} (\bibinfo {year} {2018})},\ \Eprint {http://arxiv.org/abs/1710.04663} {arXiv:1710.04663 [hep-ph]} \BibitemShut {NoStop}%
\bibitem [{\citenamefont {Haisch}\ and\ \citenamefont {Malinauskas}(2018)}]{Haisch:2017gql}%
  \BibitemOpen
  \bibfield  {author} {\bibinfo {author} {\bibfnamefont {U.}~\bibnamefont {Haisch}}\ and\ \bibinfo {author} {\bibfnamefont {A.}~\bibnamefont {Malinauskas}},\ }\href {\doibase 10.1007/JHEP03(2018)135} {\bibfield  {journal} {\bibinfo  {journal} {JHEP}\ }\textbf {\bibinfo {volume} {03}},\ \bibinfo {pages} {135} (\bibinfo {year} {2018})},\ \Eprint {http://arxiv.org/abs/1712.06599} {arXiv:1712.06599 [hep-ph]} \BibitemShut {NoStop}%
\bibitem [{\citenamefont {Fox}\ and\ \citenamefont {Weiner}(2018)}]{Fox:2017uwr}%
  \BibitemOpen
  \bibfield  {author} {\bibinfo {author} {\bibfnamefont {P.~J.}\ \bibnamefont {Fox}}\ and\ \bibinfo {author} {\bibfnamefont {N.}~\bibnamefont {Weiner}},\ }\href {\doibase 10.1007/JHEP08(2018)025} {\bibfield  {journal} {\bibinfo  {journal} {JHEP}\ }\textbf {\bibinfo {volume} {08}},\ \bibinfo {pages} {025} (\bibinfo {year} {2018})},\ \Eprint {http://arxiv.org/abs/1710.07649} {arXiv:1710.07649 [hep-ph]} \BibitemShut {NoStop}%
\bibitem [{\citenamefont {Liu}\ \emph {et~al.}(2018)\citenamefont {Liu}, \citenamefont {Liu}, \citenamefont {Wagner},\ and\ \citenamefont {Wang}}]{Liu:2018xsw}%
  \BibitemOpen
  \bibfield  {author} {\bibinfo {author} {\bibfnamefont {D.}~\bibnamefont {Liu}}, \bibinfo {author} {\bibfnamefont {J.}~\bibnamefont {Liu}}, \bibinfo {author} {\bibfnamefont {C.~E.~M.}\ \bibnamefont {Wagner}}, \ and\ \bibinfo {author} {\bibfnamefont {X.-P.}\ \bibnamefont {Wang}},\ }\href {\doibase 10.1007/JHEP06(2018)150} {\bibfield  {journal} {\bibinfo  {journal} {JHEP}\ }\textbf {\bibinfo {volume} {06}},\ \bibinfo {pages} {150} (\bibinfo {year} {2018})},\ \Eprint {http://arxiv.org/abs/1805.01476} {arXiv:1805.01476 [hep-ph]} \BibitemShut {NoStop}%
\bibitem [{\citenamefont {Wang}\ \emph {et~al.}(2018)\citenamefont {Wang}, \citenamefont {Wang}, \citenamefont {Zhu},\ and\ \citenamefont {Jie}}]{Wang:2018vxp}%
  \BibitemOpen
  \bibfield  {author} {\bibinfo {author} {\bibfnamefont {K.}~\bibnamefont {Wang}}, \bibinfo {author} {\bibfnamefont {F.}~\bibnamefont {Wang}}, \bibinfo {author} {\bibfnamefont {J.}~\bibnamefont {Zhu}}, \ and\ \bibinfo {author} {\bibfnamefont {Q.}~\bibnamefont {Jie}},\ }\href {\doibase 10.1088/1674-1137/42/10/103109} {\bibfield  {journal} {\bibinfo  {journal} {Chin. Phys. C}\ }\textbf {\bibinfo {volume} {42}},\ \bibinfo {pages} {103109} (\bibinfo {year} {2018})},\ \Eprint {http://arxiv.org/abs/1811.04435} {arXiv:1811.04435 [hep-ph]} \BibitemShut {NoStop}%
\bibitem [{\citenamefont {Liu}\ \emph {et~al.}(2019)\citenamefont {Liu}, \citenamefont {Qiao}, \citenamefont {Wang},\ and\ \citenamefont {Zhu}}]{Liu:2018ryo}%
  \BibitemOpen
  \bibfield  {author} {\bibinfo {author} {\bibfnamefont {L.}~\bibnamefont {Liu}}, \bibinfo {author} {\bibfnamefont {H.}~\bibnamefont {Qiao}}, \bibinfo {author} {\bibfnamefont {K.}~\bibnamefont {Wang}}, \ and\ \bibinfo {author} {\bibfnamefont {J.}~\bibnamefont {Zhu}},\ }\href {\doibase 10.1088/1674-1137/43/2/023104} {\bibfield  {journal} {\bibinfo  {journal} {Chin. Phys. C}\ }\textbf {\bibinfo {volume} {43}},\ \bibinfo {pages} {023104} (\bibinfo {year} {2019})},\ \Eprint {http://arxiv.org/abs/1812.00107} {arXiv:1812.00107 [hep-ph]} \BibitemShut {NoStop}%
\bibitem [{\citenamefont {Biek\"otter}\ \emph {et~al.}(2020)\citenamefont {Biek\"otter}, \citenamefont {Chakraborti},\ and\ \citenamefont {Heinemeyer}}]{Biekotter:2019kde}%
  \BibitemOpen
  \bibfield  {author} {\bibinfo {author} {\bibfnamefont {T.}~\bibnamefont {Biek\"otter}}, \bibinfo {author} {\bibfnamefont {M.}~\bibnamefont {Chakraborti}}, \ and\ \bibinfo {author} {\bibfnamefont {S.}~\bibnamefont {Heinemeyer}},\ }\href {\doibase 10.1140/epjc/s10052-019-7561-2} {\bibfield  {journal} {\bibinfo  {journal} {Eur. Phys. J. C}\ }\textbf {\bibinfo {volume} {80}},\ \bibinfo {pages} {2} (\bibinfo {year} {2020})},\ \Eprint {http://arxiv.org/abs/1903.11661} {arXiv:1903.11661 [hep-ph]} \BibitemShut {NoStop}%
\bibitem [{\citenamefont {Cline}\ and\ \citenamefont {Toma}(2019)}]{Cline:2019okt}%
  \BibitemOpen
  \bibfield  {author} {\bibinfo {author} {\bibfnamefont {J.~M.}\ \bibnamefont {Cline}}\ and\ \bibinfo {author} {\bibfnamefont {T.}~\bibnamefont {Toma}},\ }\href {\doibase 10.1103/PhysRevD.100.035023} {\bibfield  {journal} {\bibinfo  {journal} {Phys. Rev. D}\ }\textbf {\bibinfo {volume} {100}},\ \bibinfo {pages} {035023} (\bibinfo {year} {2019})},\ \Eprint {http://arxiv.org/abs/1906.02175} {arXiv:1906.02175 [hep-ph]} \BibitemShut {NoStop}%
\bibitem [{\citenamefont {Choi}\ \emph {et~al.}(2019)\citenamefont {Choi}, \citenamefont {Im}, \citenamefont {Jeong},\ and\ \citenamefont {Park}}]{Choi:2019yrv}%
  \BibitemOpen
  \bibfield  {author} {\bibinfo {author} {\bibfnamefont {K.}~\bibnamefont {Choi}}, \bibinfo {author} {\bibfnamefont {S.~H.}\ \bibnamefont {Im}}, \bibinfo {author} {\bibfnamefont {K.~S.}\ \bibnamefont {Jeong}}, \ and\ \bibinfo {author} {\bibfnamefont {C.~B.}\ \bibnamefont {Park}},\ }\href {\doibase 10.1140/epjc/s10052-019-7473-1} {\bibfield  {journal} {\bibinfo  {journal} {Eur. Phys. J. C}\ }\textbf {\bibinfo {volume} {79}},\ \bibinfo {pages} {956} (\bibinfo {year} {2019})},\ \Eprint {http://arxiv.org/abs/1906.03389} {arXiv:1906.03389 [hep-ph]} \BibitemShut {NoStop}%
\bibitem [{\citenamefont {Kundu}\ \emph {et~al.}(2020)\citenamefont {Kundu}, \citenamefont {Maharana},\ and\ \citenamefont {Mondal}}]{Kundu:2019nqo}%
  \BibitemOpen
  \bibfield  {author} {\bibinfo {author} {\bibfnamefont {A.}~\bibnamefont {Kundu}}, \bibinfo {author} {\bibfnamefont {S.}~\bibnamefont {Maharana}}, \ and\ \bibinfo {author} {\bibfnamefont {P.}~\bibnamefont {Mondal}},\ }\href {\doibase 10.1016/j.nuclphysb.2020.115057} {\bibfield  {journal} {\bibinfo  {journal} {Nucl. Phys. B}\ }\textbf {\bibinfo {volume} {955}},\ \bibinfo {pages} {115057} (\bibinfo {year} {2020})},\ \Eprint {http://arxiv.org/abs/1907.12808} {arXiv:1907.12808 [hep-ph]} \BibitemShut {NoStop}%
\bibitem [{\citenamefont {Cao}\ \emph {et~al.}(2020)\citenamefont {Cao}, \citenamefont {Jia}, \citenamefont {Yue}, \citenamefont {Zhou},\ and\ \citenamefont {Zhu}}]{Cao:2019ofo}%
  \BibitemOpen
  \bibfield  {author} {\bibinfo {author} {\bibfnamefont {J.}~\bibnamefont {Cao}}, \bibinfo {author} {\bibfnamefont {X.}~\bibnamefont {Jia}}, \bibinfo {author} {\bibfnamefont {Y.}~\bibnamefont {Yue}}, \bibinfo {author} {\bibfnamefont {H.}~\bibnamefont {Zhou}}, \ and\ \bibinfo {author} {\bibfnamefont {P.}~\bibnamefont {Zhu}},\ }\href {\doibase 10.1103/PhysRevD.101.055008} {\bibfield  {journal} {\bibinfo  {journal} {Phys. Rev. D}\ }\textbf {\bibinfo {volume} {101}},\ \bibinfo {pages} {055008} (\bibinfo {year} {2020})},\ \Eprint {http://arxiv.org/abs/1908.07206} {arXiv:1908.07206 [hep-ph]} \BibitemShut {NoStop}%
\bibitem [{\citenamefont {Biek\"otter}\ \emph {et~al.}(2021)\citenamefont {Biek\"otter}, \citenamefont {Chakraborti},\ and\ \citenamefont {Heinemeyer}}]{Biekotter:2020cjs}%
  \BibitemOpen
  \bibfield  {author} {\bibinfo {author} {\bibfnamefont {T.}~\bibnamefont {Biek\"otter}}, \bibinfo {author} {\bibfnamefont {M.}~\bibnamefont {Chakraborti}}, \ and\ \bibinfo {author} {\bibfnamefont {S.}~\bibnamefont {Heinemeyer}},\ }\href {\doibase 10.1142/S0217751X21420185} {\bibfield  {journal} {\bibinfo  {journal} {Int. J. Mod. Phys. A}\ }\textbf {\bibinfo {volume} {36}},\ \bibinfo {pages} {2142018} (\bibinfo {year} {2021})},\ \Eprint {http://arxiv.org/abs/2003.05422} {arXiv:2003.05422 [hep-ph]} \BibitemShut {NoStop}%
\bibitem [{\citenamefont {Abdelalim}\ \emph {et~al.}(2022)\citenamefont {Abdelalim}, \citenamefont {Das}, \citenamefont {Khalil},\ and\ \citenamefont {Moretti}}]{Abdelalim:2020xfk}%
  \BibitemOpen
  \bibfield  {author} {\bibinfo {author} {\bibfnamefont {A.~A.}\ \bibnamefont {Abdelalim}}, \bibinfo {author} {\bibfnamefont {B.}~\bibnamefont {Das}}, \bibinfo {author} {\bibfnamefont {S.}~\bibnamefont {Khalil}}, \ and\ \bibinfo {author} {\bibfnamefont {S.}~\bibnamefont {Moretti}},\ }\href {\doibase 10.1016/j.nuclphysb.2022.116013} {\bibfield  {journal} {\bibinfo  {journal} {Nucl. Phys. B}\ }\textbf {\bibinfo {volume} {985}},\ \bibinfo {pages} {116013} (\bibinfo {year} {2022})},\ \Eprint {http://arxiv.org/abs/2012.04952} {arXiv:2012.04952 [hep-ph]} \BibitemShut {NoStop}%
\bibitem [{\citenamefont {Heinemeyer}\ \emph {et~al.}(2022)\citenamefont {Heinemeyer}, \citenamefont {Li}, \citenamefont {Lika}, \citenamefont {Moortgat-Pick},\ and\ \citenamefont {Paasch}}]{Heinemeyer:2021msz}%
  \BibitemOpen
  \bibfield  {author} {\bibinfo {author} {\bibfnamefont {S.}~\bibnamefont {Heinemeyer}}, \bibinfo {author} {\bibfnamefont {C.}~\bibnamefont {Li}}, \bibinfo {author} {\bibfnamefont {F.}~\bibnamefont {Lika}}, \bibinfo {author} {\bibfnamefont {G.}~\bibnamefont {Moortgat-Pick}}, \ and\ \bibinfo {author} {\bibfnamefont {S.}~\bibnamefont {Paasch}},\ }\href {\doibase 10.1103/PhysRevD.106.075003} {\bibfield  {journal} {\bibinfo  {journal} {Phys. Rev. D}\ }\textbf {\bibinfo {volume} {106}},\ \bibinfo {pages} {075003} (\bibinfo {year} {2022})},\ \Eprint {http://arxiv.org/abs/2112.11958} {arXiv:2112.11958 [hep-ph]} \BibitemShut {NoStop}%
\bibitem [{\citenamefont {Biek\"otter}\ \emph {et~al.}(2022)\citenamefont {Biek\"otter}, \citenamefont {Heinemeyer},\ and\ \citenamefont {Weiglein}}]{Biekotter:2022jyr}%
  \BibitemOpen
  \bibfield  {author} {\bibinfo {author} {\bibfnamefont {T.}~\bibnamefont {Biek\"otter}}, \bibinfo {author} {\bibfnamefont {S.}~\bibnamefont {Heinemeyer}}, \ and\ \bibinfo {author} {\bibfnamefont {G.}~\bibnamefont {Weiglein}},\ }\href {\doibase 10.1007/JHEP08(2022)201} {\bibfield  {journal} {\bibinfo  {journal} {JHEP}\ }\textbf {\bibinfo {volume} {08}},\ \bibinfo {pages} {201} (\bibinfo {year} {2022})},\ \Eprint {http://arxiv.org/abs/2203.13180} {arXiv:2203.13180 [hep-ph]} \BibitemShut {NoStop}%
\bibitem [{\citenamefont {Iguro}\ \emph {et~al.}(2022)\citenamefont {Iguro}, \citenamefont {Kitahara},\ and\ \citenamefont {Omura}}]{Iguro:2022dok}%
  \BibitemOpen
  \bibfield  {author} {\bibinfo {author} {\bibfnamefont {S.}~\bibnamefont {Iguro}}, \bibinfo {author} {\bibfnamefont {T.}~\bibnamefont {Kitahara}}, \ and\ \bibinfo {author} {\bibfnamefont {Y.}~\bibnamefont {Omura}},\ }\href {\doibase 10.1140/epjc/s10052-022-11028-y} {\bibfield  {journal} {\bibinfo  {journal} {Eur. Phys. J. C}\ }\textbf {\bibinfo {volume} {82}},\ \bibinfo {pages} {1053} (\bibinfo {year} {2022})},\ \Eprint {http://arxiv.org/abs/2205.03187} {arXiv:2205.03187 [hep-ph]} \BibitemShut {NoStop}%
\bibitem [{\citenamefont {Li}\ \emph {et~al.}(2022)\citenamefont {Li}, \citenamefont {Zhu}, \citenamefont {Wang}, \citenamefont {Ma}, \citenamefont {Tian},\ and\ \citenamefont {Qiao}}]{Li:2022etb}%
  \BibitemOpen
  \bibfield  {author} {\bibinfo {author} {\bibfnamefont {W.}~\bibnamefont {Li}}, \bibinfo {author} {\bibfnamefont {J.}~\bibnamefont {Zhu}}, \bibinfo {author} {\bibfnamefont {K.}~\bibnamefont {Wang}}, \bibinfo {author} {\bibfnamefont {S.}~\bibnamefont {Ma}}, \bibinfo {author} {\bibfnamefont {P.}~\bibnamefont {Tian}}, \ and\ \bibinfo {author} {\bibfnamefont {H.}~\bibnamefont {Qiao}},\ }\href@noop {} {\  (\bibinfo {year} {2022})},\ \Eprint {http://arxiv.org/abs/2212.11739} {arXiv:2212.11739 [hep-ph]} \BibitemShut {NoStop}%
\bibitem [{\citenamefont {Biek\"otter}\ \emph {et~al.}(2023)\citenamefont {Biek\"otter}, \citenamefont {Heinemeyer},\ and\ \citenamefont {Weiglein}}]{Biekotter:2023jld}%
  \BibitemOpen
  \bibfield  {author} {\bibinfo {author} {\bibfnamefont {T.}~\bibnamefont {Biek\"otter}}, \bibinfo {author} {\bibfnamefont {S.}~\bibnamefont {Heinemeyer}}, \ and\ \bibinfo {author} {\bibfnamefont {G.}~\bibnamefont {Weiglein}},\ }\href@noop {} {\  (\bibinfo {year} {2023})},\ \Eprint {http://arxiv.org/abs/2303.12018} {arXiv:2303.12018 [hep-ph]} \BibitemShut {NoStop}%
\bibitem [{\citenamefont {Bonilla}\ \emph {et~al.}(2023)\citenamefont {Bonilla}, \citenamefont {Carcamo~Hernandez}, \citenamefont {Kovalenko}, \citenamefont {Lee}, \citenamefont {Pasechnik},\ and\ \citenamefont {Schmidt}}]{Bonilla:2023wok}%
  \BibitemOpen
  \bibfield  {author} {\bibinfo {author} {\bibfnamefont {C.}~\bibnamefont {Bonilla}}, \bibinfo {author} {\bibfnamefont {A.~E.}\ \bibnamefont {Carcamo~Hernandez}}, \bibinfo {author} {\bibfnamefont {S.}~\bibnamefont {Kovalenko}}, \bibinfo {author} {\bibfnamefont {H.}~\bibnamefont {Lee}}, \bibinfo {author} {\bibfnamefont {R.}~\bibnamefont {Pasechnik}}, \ and\ \bibinfo {author} {\bibfnamefont {I.}~\bibnamefont {Schmidt}},\ }\href@noop {} {\  (\bibinfo {year} {2023})},\ \Eprint {http://arxiv.org/abs/2305.11967} {arXiv:2305.11967 [hep-ph]} \BibitemShut {NoStop}%
\bibitem [{\citenamefont {Azevedo}\ \emph {et~al.}(2023)\citenamefont {Azevedo}, \citenamefont {Biek\"otter},\ and\ \citenamefont {Ferreira}}]{Azevedo:2023zkg}%
  \BibitemOpen
  \bibfield  {author} {\bibinfo {author} {\bibfnamefont {D.}~\bibnamefont {Azevedo}}, \bibinfo {author} {\bibfnamefont {T.}~\bibnamefont {Biek\"otter}}, \ and\ \bibinfo {author} {\bibfnamefont {P.~M.}\ \bibnamefont {Ferreira}},\ }\href@noop {} {\  (\bibinfo {year} {2023})},\ \Eprint {http://arxiv.org/abs/2305.19716} {arXiv:2305.19716 [hep-ph]} \BibitemShut {NoStop}%
\bibitem [{\citenamefont {Escribano}\ \emph {et~al.}(2023)\citenamefont {Escribano}, \citenamefont {Lozano},\ and\ \citenamefont {Vicente}}]{Escribano:2023hxj}%
  \BibitemOpen
  \bibfield  {author} {\bibinfo {author} {\bibfnamefont {P.}~\bibnamefont {Escribano}}, \bibinfo {author} {\bibfnamefont {V.~M.}\ \bibnamefont {Lozano}}, \ and\ \bibinfo {author} {\bibfnamefont {A.}~\bibnamefont {Vicente}},\ }\href@noop {} {\  (\bibinfo {year} {2023})},\ \Eprint {http://arxiv.org/abs/2306.03735} {arXiv:2306.03735 [hep-ph]} \BibitemShut {NoStop}%
\bibitem [{\citenamefont {Belyaev}\ \emph {et~al.}(2023)\citenamefont {Belyaev}, \citenamefont {Benbrik}, \citenamefont {Boukidi}, \citenamefont {Chakraborti}, \citenamefont {Moretti},\ and\ \citenamefont {Semlali}}]{Belyaev:2023xnv}%
  \BibitemOpen
  \bibfield  {author} {\bibinfo {author} {\bibfnamefont {A.}~\bibnamefont {Belyaev}}, \bibinfo {author} {\bibfnamefont {R.}~\bibnamefont {Benbrik}}, \bibinfo {author} {\bibfnamefont {M.}~\bibnamefont {Boukidi}}, \bibinfo {author} {\bibfnamefont {M.}~\bibnamefont {Chakraborti}}, \bibinfo {author} {\bibfnamefont {S.}~\bibnamefont {Moretti}}, \ and\ \bibinfo {author} {\bibfnamefont {S.}~\bibnamefont {Semlali}},\ }\href@noop {} {\  (\bibinfo {year} {2023})},\ \Eprint {http://arxiv.org/abs/2306.09029} {arXiv:2306.09029 [hep-ph]} \BibitemShut {NoStop}%
\bibitem [{\citenamefont {Ashanujjaman}\ \emph {et~al.}(2023)\citenamefont {Ashanujjaman}, \citenamefont {Banik}, \citenamefont {Coloretti}, \citenamefont {Crivellin}, \citenamefont {Mellado},\ and\ \citenamefont {Mulaudzi}}]{Ashanujjaman:2023etj}%
  \BibitemOpen
  \bibfield  {author} {\bibinfo {author} {\bibfnamefont {S.}~\bibnamefont {Ashanujjaman}}, \bibinfo {author} {\bibfnamefont {S.}~\bibnamefont {Banik}}, \bibinfo {author} {\bibfnamefont {G.}~\bibnamefont {Coloretti}}, \bibinfo {author} {\bibfnamefont {A.}~\bibnamefont {Crivellin}}, \bibinfo {author} {\bibfnamefont {B.}~\bibnamefont {Mellado}}, \ and\ \bibinfo {author} {\bibfnamefont {A.-T.}\ \bibnamefont {Mulaudzi}},\ }\href {\doibase 10.1103/PhysRevD.108.L091704} {\bibfield  {journal} {\bibinfo  {journal} {Phys. Rev. D}\ }\textbf {\bibinfo {volume} {108}},\ \bibinfo {pages} {L091704} (\bibinfo {year} {2023})},\ \Eprint {http://arxiv.org/abs/2306.15722} {arXiv:2306.15722 [hep-ph]} \BibitemShut {NoStop}%
\bibitem [{\citenamefont {Baek}\ \emph {et~al.}(2024)\citenamefont {Baek}, \citenamefont {Ko}, \citenamefont {Omura},\ and\ \citenamefont {Yu}}]{Baek:2024cco}%
  \BibitemOpen
  \bibfield  {author} {\bibinfo {author} {\bibfnamefont {S.}~\bibnamefont {Baek}}, \bibinfo {author} {\bibfnamefont {P.}~\bibnamefont {Ko}}, \bibinfo {author} {\bibfnamefont {Y.}~\bibnamefont {Omura}}, \ and\ \bibinfo {author} {\bibfnamefont {C.}~\bibnamefont {Yu}},\ }\href@noop {} {\  (\bibinfo {year} {2024})},\ \Eprint {http://arxiv.org/abs/2412.02178} {arXiv:2412.02178 [hep-ph]} \BibitemShut {NoStop}%
\bibitem [{\citenamefont {Du}\ \emph {et~al.}(2025)\citenamefont {Du}, \citenamefont {Liu},\ and\ \citenamefont {Chang}}]{Du:2025eop}%
  \BibitemOpen
  \bibfield  {author} {\bibinfo {author} {\bibfnamefont {X.}~\bibnamefont {Du}}, \bibinfo {author} {\bibfnamefont {H.}~\bibnamefont {Liu}}, \ and\ \bibinfo {author} {\bibfnamefont {Q.}~\bibnamefont {Chang}},\ }\href@noop {} {\  (\bibinfo {year} {2025})},\ \Eprint {http://arxiv.org/abs/2502.06444} {arXiv:2502.06444 [hep-ph]} \BibitemShut {NoStop}%
\bibitem [{\citenamefont {Yaser~Ayazi}\ \emph {et~al.}(2024)\citenamefont {Yaser~Ayazi}, \citenamefont {Hosseini}, \citenamefont {Paktinat~Mehdiabadi},\ and\ \citenamefont {Rouzbehi}}]{YaserAyazi:2024hpj}%
  \BibitemOpen
  \bibfield  {author} {\bibinfo {author} {\bibfnamefont {S.}~\bibnamefont {Yaser~Ayazi}}, \bibinfo {author} {\bibfnamefont {M.}~\bibnamefont {Hosseini}}, \bibinfo {author} {\bibfnamefont {S.}~\bibnamefont {Paktinat~Mehdiabadi}}, \ and\ \bibinfo {author} {\bibfnamefont {R.}~\bibnamefont {Rouzbehi}},\ }\href {\doibase 10.1103/PhysRevD.110.055004} {\bibfield  {journal} {\bibinfo  {journal} {Phys. Rev. D}\ }\textbf {\bibinfo {volume} {110}},\ \bibinfo {pages} {055004} (\bibinfo {year} {2024})},\ \Eprint {http://arxiv.org/abs/2405.01132} {arXiv:2405.01132 [hep-ph]} \BibitemShut {NoStop}%
\bibitem [{\citenamefont {Borah}\ \emph {et~al.}(2024)\citenamefont {Borah}, \citenamefont {Mahapatra}, \citenamefont {Paul},\ and\ \citenamefont {Sahu}}]{Borah:2023hqw}%
  \BibitemOpen
  \bibfield  {author} {\bibinfo {author} {\bibfnamefont {D.}~\bibnamefont {Borah}}, \bibinfo {author} {\bibfnamefont {S.}~\bibnamefont {Mahapatra}}, \bibinfo {author} {\bibfnamefont {P.~K.}\ \bibnamefont {Paul}}, \ and\ \bibinfo {author} {\bibfnamefont {N.}~\bibnamefont {Sahu}},\ }\href {\doibase 10.1103/PhysRevD.109.055021} {\bibfield  {journal} {\bibinfo  {journal} {Phys. Rev. D}\ }\textbf {\bibinfo {volume} {109}},\ \bibinfo {pages} {055021} (\bibinfo {year} {2024})},\ \Eprint {http://arxiv.org/abs/2310.11953} {arXiv:2310.11953 [hep-ph]} \BibitemShut {NoStop}%
\bibitem [{\citenamefont {Bhattacharya}\ \emph {et~al.}(2023)\citenamefont {Bhattacharya}, \citenamefont {Coloretti}, \citenamefont {Crivellin}, \citenamefont {Dahbi}, \citenamefont {Fang}, \citenamefont {Kumar},\ and\ \citenamefont {Mellado}}]{Bhattacharya:2023lmu}%
  \BibitemOpen
  \bibfield  {author} {\bibinfo {author} {\bibfnamefont {S.}~\bibnamefont {Bhattacharya}}, \bibinfo {author} {\bibfnamefont {G.}~\bibnamefont {Coloretti}}, \bibinfo {author} {\bibfnamefont {A.}~\bibnamefont {Crivellin}}, \bibinfo {author} {\bibfnamefont {S.-E.}\ \bibnamefont {Dahbi}}, \bibinfo {author} {\bibfnamefont {Y.}~\bibnamefont {Fang}}, \bibinfo {author} {\bibfnamefont {M.}~\bibnamefont {Kumar}}, \ and\ \bibinfo {author} {\bibfnamefont {B.}~\bibnamefont {Mellado}},\ }\href@noop {} {\  (\bibinfo {year} {2023})},\ \Eprint {http://arxiv.org/abs/2306.17209} {arXiv:2306.17209 [hep-ph]} \BibitemShut {NoStop}%
\bibitem [{\citenamefont {Coloretti}\ \emph {et~al.}(2024)\citenamefont {Coloretti}, \citenamefont {Crivellin},\ and\ \citenamefont {Mellado}}]{Coloretti:2023yyq}%
  \BibitemOpen
  \bibfield  {author} {\bibinfo {author} {\bibfnamefont {G.}~\bibnamefont {Coloretti}}, \bibinfo {author} {\bibfnamefont {A.}~\bibnamefont {Crivellin}}, \ and\ \bibinfo {author} {\bibfnamefont {B.}~\bibnamefont {Mellado}},\ }\href {\doibase 10.1103/PhysRevD.110.073001} {\bibfield  {journal} {\bibinfo  {journal} {Phys. Rev. D}\ }\textbf {\bibinfo {volume} {110}},\ \bibinfo {pages} {073001} (\bibinfo {year} {2024})},\ \Eprint {http://arxiv.org/abs/2312.17314} {arXiv:2312.17314 [hep-ph]} \BibitemShut {NoStop}%
\bibitem [{\citenamefont {Banik}\ \emph {et~al.}(2025)\citenamefont {Banik}, \citenamefont {Coloretti}, \citenamefont {Crivellin},\ and\ \citenamefont {Mellado}}]{Banik:2023vxa}%
  \BibitemOpen
  \bibfield  {author} {\bibinfo {author} {\bibfnamefont {S.}~\bibnamefont {Banik}}, \bibinfo {author} {\bibfnamefont {G.}~\bibnamefont {Coloretti}}, \bibinfo {author} {\bibfnamefont {A.}~\bibnamefont {Crivellin}}, \ and\ \bibinfo {author} {\bibfnamefont {B.}~\bibnamefont {Mellado}},\ }\href {\doibase 10.1007/JHEP01(2025)155} {\bibfield  {journal} {\bibinfo  {journal} {JHEP}\ }\textbf {\bibinfo {volume} {01}},\ \bibinfo {pages} {155} (\bibinfo {year} {2025})},\ \Eprint {http://arxiv.org/abs/2308.07953} {arXiv:2308.07953 [hep-ph]} \BibitemShut {NoStop}%
\bibitem [{\citenamefont {Ashanujjaman}\ \emph {et~al.}(2025)\citenamefont {Ashanujjaman}, \citenamefont {Banik}, \citenamefont {Coloretti}, \citenamefont {Crivellin}, \citenamefont {Maharathy},\ and\ \citenamefont {Mellado}}]{Ashanujjaman:2024lnr}%
  \BibitemOpen
  \bibfield  {author} {\bibinfo {author} {\bibfnamefont {S.}~\bibnamefont {Ashanujjaman}}, \bibinfo {author} {\bibfnamefont {S.}~\bibnamefont {Banik}}, \bibinfo {author} {\bibfnamefont {G.}~\bibnamefont {Coloretti}}, \bibinfo {author} {\bibfnamefont {A.}~\bibnamefont {Crivellin}}, \bibinfo {author} {\bibfnamefont {S.~P.}\ \bibnamefont {Maharathy}}, \ and\ \bibinfo {author} {\bibfnamefont {B.}~\bibnamefont {Mellado}},\ }\href {\doibase 10.1007/JHEP04(2025)003} {\bibfield  {journal} {\bibinfo  {journal} {JHEP}\ }\textbf {\bibinfo {volume} {04}},\ \bibinfo {pages} {003} (\bibinfo {year} {2025})},\ \Eprint {http://arxiv.org/abs/2411.18618} {arXiv:2411.18618 [hep-ph]} \BibitemShut {NoStop}%
\bibitem [{\citenamefont {Dev}\ \emph {et~al.}(2024)\citenamefont {Dev}, \citenamefont {Mohapatra},\ and\ \citenamefont {Zhang}}]{Dev:2023kzu}%
  \BibitemOpen
  \bibfield  {author} {\bibinfo {author} {\bibfnamefont {P.~S.~B.}\ \bibnamefont {Dev}}, \bibinfo {author} {\bibfnamefont {R.~N.}\ \bibnamefont {Mohapatra}}, \ and\ \bibinfo {author} {\bibfnamefont {Y.}~\bibnamefont {Zhang}},\ }\href {\doibase 10.1016/j.physletb.2024.138481} {\bibfield  {journal} {\bibinfo  {journal} {Phys. Lett. B}\ }\textbf {\bibinfo {volume} {849}},\ \bibinfo {pages} {138481} (\bibinfo {year} {2024})},\ \Eprint {http://arxiv.org/abs/2312.17733} {arXiv:2312.17733 [hep-ph]} \BibitemShut {NoStop}%
\bibitem [{\citenamefont {Mondal}\ \emph {et~al.}(2024)\citenamefont {Mondal}, \citenamefont {Moretti},\ and\ \citenamefont {Sanyal}}]{Mondal:2024obd}%
  \BibitemOpen
  \bibfield  {author} {\bibinfo {author} {\bibfnamefont {T.}~\bibnamefont {Mondal}}, \bibinfo {author} {\bibfnamefont {S.}~\bibnamefont {Moretti}}, \ and\ \bibinfo {author} {\bibfnamefont {P.}~\bibnamefont {Sanyal}},\ }\href@noop {} {\  (\bibinfo {year} {2024})},\ \Eprint {http://arxiv.org/abs/2412.00474} {arXiv:2412.00474 [hep-ph]} \BibitemShut {NoStop}%
\bibitem [{\citenamefont {Cao}\ \emph {et~al.}(2024{\natexlab{a}})\citenamefont {Cao}, \citenamefont {Jia}, \citenamefont {Lian},\ and\ \citenamefont {Meng}}]{Cao:2023gkc}%
  \BibitemOpen
  \bibfield  {author} {\bibinfo {author} {\bibfnamefont {J.}~\bibnamefont {Cao}}, \bibinfo {author} {\bibfnamefont {X.}~\bibnamefont {Jia}}, \bibinfo {author} {\bibfnamefont {J.}~\bibnamefont {Lian}}, \ and\ \bibinfo {author} {\bibfnamefont {L.}~\bibnamefont {Meng}},\ }\href {\doibase 10.1103/PhysRevD.109.075001} {\bibfield  {journal} {\bibinfo  {journal} {Phys. Rev. D}\ }\textbf {\bibinfo {volume} {109}},\ \bibinfo {pages} {075001} (\bibinfo {year} {2024}{\natexlab{a}})},\ \Eprint {http://arxiv.org/abs/2310.08436} {arXiv:2310.08436 [hep-ph]} \BibitemShut {NoStop}%
\bibitem [{\citenamefont {Ellwanger}\ and\ \citenamefont {Hugonie}(2023)}]{Ellwanger:2023zjc}%
  \BibitemOpen
  \bibfield  {author} {\bibinfo {author} {\bibfnamefont {U.}~\bibnamefont {Ellwanger}}\ and\ \bibinfo {author} {\bibfnamefont {C.}~\bibnamefont {Hugonie}},\ }\href {\doibase 10.1140/epjc/s10052-023-12315-y} {\bibfield  {journal} {\bibinfo  {journal} {Eur. Phys. J. C}\ }\textbf {\bibinfo {volume} {83}},\ \bibinfo {pages} {1138} (\bibinfo {year} {2023})},\ \Eprint {http://arxiv.org/abs/2309.07838} {arXiv:2309.07838 [hep-ph]} \BibitemShut {NoStop}%
\bibitem [{\citenamefont {Lian}(2024)}]{Lian:2024smg}%
  \BibitemOpen
  \bibfield  {author} {\bibinfo {author} {\bibfnamefont {J.}~\bibnamefont {Lian}},\ }\href {\doibase 10.1103/PhysRevD.110.115018} {\bibfield  {journal} {\bibinfo  {journal} {Phys. Rev. D}\ }\textbf {\bibinfo {volume} {110}},\ \bibinfo {pages} {115018} (\bibinfo {year} {2024})},\ \Eprint {http://arxiv.org/abs/2406.10969} {arXiv:2406.10969 [hep-ph]} \BibitemShut {NoStop}%
\bibitem [{\citenamefont {Ellwanger}\ \emph {et~al.}(2024)\citenamefont {Ellwanger}, \citenamefont {Hugonie}, \citenamefont {King},\ and\ \citenamefont {Moretti}}]{Ellwanger:2024vvs}%
  \BibitemOpen
  \bibfield  {author} {\bibinfo {author} {\bibfnamefont {U.}~\bibnamefont {Ellwanger}}, \bibinfo {author} {\bibfnamefont {C.}~\bibnamefont {Hugonie}}, \bibinfo {author} {\bibfnamefont {S.~F.}\ \bibnamefont {King}}, \ and\ \bibinfo {author} {\bibfnamefont {S.}~\bibnamefont {Moretti}},\ }\href {\doibase 10.1140/epjc/s10052-024-13129-2} {\bibfield  {journal} {\bibinfo  {journal} {Eur. Phys. J. C}\ }\textbf {\bibinfo {volume} {84}},\ \bibinfo {pages} {788} (\bibinfo {year} {2024})},\ \Eprint {http://arxiv.org/abs/2404.19338} {arXiv:2404.19338 [hep-ph]} \BibitemShut {NoStop}%
\bibitem [{\citenamefont {Ellwanger}\ and\ \citenamefont {Hugonie}(2024)}]{Ellwanger:2024txc}%
  \BibitemOpen
  \bibfield  {author} {\bibinfo {author} {\bibfnamefont {U.}~\bibnamefont {Ellwanger}}\ and\ \bibinfo {author} {\bibfnamefont {C.}~\bibnamefont {Hugonie}},\ }\href {\doibase 10.1140/epjc/s10052-024-12886-4} {\bibfield  {journal} {\bibinfo  {journal} {Eur. Phys. J. C}\ }\textbf {\bibinfo {volume} {84}},\ \bibinfo {pages} {526} (\bibinfo {year} {2024})},\ \Eprint {http://arxiv.org/abs/2403.16884} {arXiv:2403.16884 [hep-ph]} \BibitemShut {NoStop}%
\bibitem [{\citenamefont {Cao}\ \emph {et~al.}(2024{\natexlab{b}})\citenamefont {Cao}, \citenamefont {Jia},\ and\ \citenamefont {Lian}}]{Cao:2024axg}%
  \BibitemOpen
  \bibfield  {author} {\bibinfo {author} {\bibfnamefont {J.}~\bibnamefont {Cao}}, \bibinfo {author} {\bibfnamefont {X.}~\bibnamefont {Jia}}, \ and\ \bibinfo {author} {\bibfnamefont {J.}~\bibnamefont {Lian}},\ }\href {\doibase 10.1103/PhysRevD.110.115039} {\bibfield  {journal} {\bibinfo  {journal} {Phys. Rev. D}\ }\textbf {\bibinfo {volume} {110}},\ \bibinfo {pages} {115039} (\bibinfo {year} {2024}{\natexlab{b}})},\ \Eprint {http://arxiv.org/abs/2402.15847} {arXiv:2402.15847 [hep-ph]} \BibitemShut {NoStop}%
\bibitem [{\citenamefont {Liu}\ \emph {et~al.}(2024)\citenamefont {Liu}, \citenamefont {Zhou}, \citenamefont {Zheng}, \citenamefont {Ma}, \citenamefont {Feng},\ and\ \citenamefont {Zhang}}]{Liu:2024cbr}%
  \BibitemOpen
  \bibfield  {author} {\bibinfo {author} {\bibfnamefont {C.-X.}\ \bibnamefont {Liu}}, \bibinfo {author} {\bibfnamefont {Y.}~\bibnamefont {Zhou}}, \bibinfo {author} {\bibfnamefont {X.-Y.}\ \bibnamefont {Zheng}}, \bibinfo {author} {\bibfnamefont {J.}~\bibnamefont {Ma}}, \bibinfo {author} {\bibfnamefont {T.-F.}\ \bibnamefont {Feng}}, \ and\ \bibinfo {author} {\bibfnamefont {H.-B.}\ \bibnamefont {Zhang}},\ }\href {\doibase 10.1103/PhysRevD.109.056001} {\bibfield  {journal} {\bibinfo  {journal} {Phys. Rev. D}\ }\textbf {\bibinfo {volume} {109}},\ \bibinfo {pages} {056001} (\bibinfo {year} {2024})},\ \Eprint {http://arxiv.org/abs/2402.00727} {arXiv:2402.00727 [hep-ph]} \BibitemShut {NoStop}%
\bibitem [{\citenamefont {Edsjo}\ and\ \citenamefont {Gondolo}(1997)}]{Edsjo:1997bg}%
  \BibitemOpen
  \bibfield  {author} {\bibinfo {author} {\bibfnamefont {J.}~\bibnamefont {Edsjo}}\ and\ \bibinfo {author} {\bibfnamefont {P.}~\bibnamefont {Gondolo}},\ }\href {\doibase 10.1103/PhysRevD.56.1879} {\bibfield  {journal} {\bibinfo  {journal} {Phys. Rev. D}\ }\textbf {\bibinfo {volume} {56}},\ \bibinfo {pages} {1879} (\bibinfo {year} {1997})},\ \Eprint {http://arxiv.org/abs/hep-ph/9704361} {arXiv:hep-ph/9704361} \BibitemShut {NoStop}%
\bibitem [{\citenamefont {Delgado}\ \emph {et~al.}(2015)\citenamefont {Delgado}, \citenamefont {Garcia-Pepin}, \citenamefont {Ostdiek},\ and\ \citenamefont {Quiros}}]{Delgado:2015aha}%
  \BibitemOpen
  \bibfield  {author} {\bibinfo {author} {\bibfnamefont {A.}~\bibnamefont {Delgado}}, \bibinfo {author} {\bibfnamefont {M.}~\bibnamefont {Garcia-Pepin}}, \bibinfo {author} {\bibfnamefont {B.}~\bibnamefont {Ostdiek}}, \ and\ \bibinfo {author} {\bibfnamefont {M.}~\bibnamefont {Quiros}},\ }\href {\doibase 10.1103/PhysRevD.92.015011} {\bibfield  {journal} {\bibinfo  {journal} {Phys. Rev. D}\ }\textbf {\bibinfo {volume} {92}},\ \bibinfo {pages} {015011} (\bibinfo {year} {2015})},\ \Eprint {http://arxiv.org/abs/1504.02486} {arXiv:1504.02486 [hep-ph]} \BibitemShut {NoStop}%
\bibitem [{\citenamefont {Djouadi}(2008)}]{Djouadi:2005gj}%
  \BibitemOpen
  \bibfield  {author} {\bibinfo {author} {\bibfnamefont {A.}~\bibnamefont {Djouadi}},\ }\href {\doibase 10.1016/j.physrep.2007.10.005} {\bibfield  {journal} {\bibinfo  {journal} {Phys. Rept.}\ }\textbf {\bibinfo {volume} {459}},\ \bibinfo {pages} {1} (\bibinfo {year} {2008})},\ \Eprint {http://arxiv.org/abs/hep-ph/0503173} {arXiv:hep-ph/0503173} \BibitemShut {NoStop}%
\bibitem [{\citenamefont {Abdughani}\ \emph {et~al.}(2018)\citenamefont {Abdughani}, \citenamefont {Wu},\ and\ \citenamefont {Yang}}]{Abdughani:2017dqs}%
  \BibitemOpen
  \bibfield  {author} {\bibinfo {author} {\bibfnamefont {M.}~\bibnamefont {Abdughani}}, \bibinfo {author} {\bibfnamefont {L.}~\bibnamefont {Wu}}, \ and\ \bibinfo {author} {\bibfnamefont {J.~M.}\ \bibnamefont {Yang}},\ }\href {\doibase 10.1140/epjc/s10052-017-5485-2} {\bibfield  {journal} {\bibinfo  {journal} {Eur. Phys. J. C}\ }\textbf {\bibinfo {volume} {78}},\ \bibinfo {pages} {4} (\bibinfo {year} {2018})},\ \Eprint {http://arxiv.org/abs/1705.09164} {arXiv:1705.09164 [hep-ph]} \BibitemShut {NoStop}%
\bibitem [{\citenamefont {Duan}\ \emph {et~al.}(2018{\natexlab{b}})\citenamefont {Duan}, \citenamefont {Hikasa}, \citenamefont {Ren}, \citenamefont {Wu},\ and\ \citenamefont {Yang}}]{Duan:2018rls}%
  \BibitemOpen
  \bibfield  {author} {\bibinfo {author} {\bibfnamefont {G.~H.}\ \bibnamefont {Duan}}, \bibinfo {author} {\bibfnamefont {K.-I.}\ \bibnamefont {Hikasa}}, \bibinfo {author} {\bibfnamefont {J.}~\bibnamefont {Ren}}, \bibinfo {author} {\bibfnamefont {L.}~\bibnamefont {Wu}}, \ and\ \bibinfo {author} {\bibfnamefont {J.~M.}\ \bibnamefont {Yang}},\ }\href {\doibase 10.1103/PhysRevD.98.015010} {\bibfield  {journal} {\bibinfo  {journal} {Phys. Rev. D}\ }\textbf {\bibinfo {volume} {98}},\ \bibinfo {pages} {015010} (\bibinfo {year} {2018}{\natexlab{b}})},\ \Eprint {http://arxiv.org/abs/1804.05238} {arXiv:1804.05238 [hep-ph]} \BibitemShut {NoStop}%
\bibitem [{\citenamefont {Abdughani}\ and\ \citenamefont {Wu}(2020)}]{Abdughani:2019wss}%
  \BibitemOpen
  \bibfield  {author} {\bibinfo {author} {\bibfnamefont {M.}~\bibnamefont {Abdughani}}\ and\ \bibinfo {author} {\bibfnamefont {L.}~\bibnamefont {Wu}},\ }\href {\doibase 10.1140/epjc/s10052-020-7793-1} {\bibfield  {journal} {\bibinfo  {journal} {Eur. Phys. J. C}\ }\textbf {\bibinfo {volume} {80}},\ \bibinfo {pages} {233} (\bibinfo {year} {2020})},\ \Eprint {http://arxiv.org/abs/1908.11350} {arXiv:1908.11350 [hep-ph]} \BibitemShut {NoStop}%
\bibitem [{\citenamefont {Duan}\ \emph {et~al.}(2019)\citenamefont {Duan}, \citenamefont {Han}, \citenamefont {Peng}, \citenamefont {Wu},\ and\ \citenamefont {Yang}}]{Duan:2018cgb}%
  \BibitemOpen
  \bibfield  {author} {\bibinfo {author} {\bibfnamefont {G.~H.}\ \bibnamefont {Duan}}, \bibinfo {author} {\bibfnamefont {C.}~\bibnamefont {Han}}, \bibinfo {author} {\bibfnamefont {B.}~\bibnamefont {Peng}}, \bibinfo {author} {\bibfnamefont {L.}~\bibnamefont {Wu}}, \ and\ \bibinfo {author} {\bibfnamefont {J.~M.}\ \bibnamefont {Yang}},\ }\href {\doibase 10.1016/j.physletb.2018.12.001} {\bibfield  {journal} {\bibinfo  {journal} {Phys. Lett. B}\ }\textbf {\bibinfo {volume} {788}},\ \bibinfo {pages} {475} (\bibinfo {year} {2019})},\ \Eprint {http://arxiv.org/abs/1809.10061} {arXiv:1809.10061 [hep-ph]} \BibitemShut {NoStop}%
\bibitem [{\citenamefont {Arkani-Hamed}\ \emph {et~al.}(2006)\citenamefont {Arkani-Hamed}, \citenamefont {Delgado},\ and\ \citenamefont {Giudice}}]{Arkani-Hamed:2006wnf}%
  \BibitemOpen
  \bibfield  {author} {\bibinfo {author} {\bibfnamefont {N.}~\bibnamefont {Arkani-Hamed}}, \bibinfo {author} {\bibfnamefont {A.}~\bibnamefont {Delgado}}, \ and\ \bibinfo {author} {\bibfnamefont {G.~F.}\ \bibnamefont {Giudice}},\ }\href {\doibase 10.1016/j.nuclphysb.2006.02.010} {\bibfield  {journal} {\bibinfo  {journal} {Nucl. Phys. B}\ }\textbf {\bibinfo {volume} {741}},\ \bibinfo {pages} {108} (\bibinfo {year} {2006})},\ \Eprint {http://arxiv.org/abs/hep-ph/0601041} {arXiv:hep-ph/0601041} \BibitemShut {NoStop}%
\bibitem [{\citenamefont {Yang}\ \emph {et~al.}(2024)\citenamefont {Yang}, \citenamefont {Yang}, \citenamefont {Zhao}, \citenamefont {Wu},\ and\ \citenamefont {Feng}}]{Yang:2024uoq}%
  \BibitemOpen
  \bibfield  {author} {\bibinfo {author} {\bibfnamefont {Z.-J.}\ \bibnamefont {Yang}}, \bibinfo {author} {\bibfnamefont {J.-L.}\ \bibnamefont {Yang}}, \bibinfo {author} {\bibfnamefont {S.-M.}\ \bibnamefont {Zhao}}, \bibinfo {author} {\bibfnamefont {X.-G.}\ \bibnamefont {Wu}}, \ and\ \bibinfo {author} {\bibfnamefont {T.-F.}\ \bibnamefont {Feng}},\ }\href {\doibase 10.1140/epjc/s10052-024-13592-x} {\bibfield  {journal} {\bibinfo  {journal} {Eur. Phys. J. C}\ }\textbf {\bibinfo {volume} {84}},\ \bibinfo {pages} {1216} (\bibinfo {year} {2024})},\ \Eprint {http://arxiv.org/abs/2410.13659} {arXiv:2410.13659 [hep-ph]} \BibitemShut {NoStop}%
\bibitem [{\citenamefont {Cheung}\ \emph {et~al.}(2013)\citenamefont {Cheung}, \citenamefont {Hall}, \citenamefont {Pinner},\ and\ \citenamefont {Ruderman}}]{Cheung:2012qy}%
  \BibitemOpen
  \bibfield  {author} {\bibinfo {author} {\bibfnamefont {C.}~\bibnamefont {Cheung}}, \bibinfo {author} {\bibfnamefont {L.~J.}\ \bibnamefont {Hall}}, \bibinfo {author} {\bibfnamefont {D.}~\bibnamefont {Pinner}}, \ and\ \bibinfo {author} {\bibfnamefont {J.~T.}\ \bibnamefont {Ruderman}},\ }\href {\doibase 10.1007/JHEP05(2013)100} {\bibfield  {journal} {\bibinfo  {journal} {JHEP}\ }\textbf {\bibinfo {volume} {05}},\ \bibinfo {pages} {100} (\bibinfo {year} {2013})},\ \Eprint {http://arxiv.org/abs/1211.4873} {arXiv:1211.4873 [hep-ph]} \BibitemShut {NoStop}%
\bibitem [{\citenamefont {Basak}\ and\ \citenamefont {Mohanty}(2013)}]{Basak:2013eba}%
  \BibitemOpen
  \bibfield  {author} {\bibinfo {author} {\bibfnamefont {T.}~\bibnamefont {Basak}}\ and\ \bibinfo {author} {\bibfnamefont {S.}~\bibnamefont {Mohanty}},\ }\href {\doibase 10.1007/JHEP08(2013)020} {\bibfield  {journal} {\bibinfo  {journal} {JHEP}\ }\textbf {\bibinfo {volume} {08}},\ \bibinfo {pages} {020} (\bibinfo {year} {2013})},\ \Eprint {http://arxiv.org/abs/1304.6856} {arXiv:1304.6856 [hep-ph]} \BibitemShut {NoStop}%
\bibitem [{\citenamefont {Ellis}\ \emph {et~al.}(2000)\citenamefont {Ellis}, \citenamefont {Ferstl},\ and\ \citenamefont {Olive}}]{Ellis:2000ds}%
  \BibitemOpen
  \bibfield  {author} {\bibinfo {author} {\bibfnamefont {J.~R.}\ \bibnamefont {Ellis}}, \bibinfo {author} {\bibfnamefont {A.}~\bibnamefont {Ferstl}}, \ and\ \bibinfo {author} {\bibfnamefont {K.~A.}\ \bibnamefont {Olive}},\ }\href {\doibase 10.1016/S0370-2693(00)00459-7} {\bibfield  {journal} {\bibinfo  {journal} {Phys. Lett. B}\ }\textbf {\bibinfo {volume} {481}},\ \bibinfo {pages} {304} (\bibinfo {year} {2000})},\ \Eprint {http://arxiv.org/abs/hep-ph/0001005} {arXiv:hep-ph/0001005} \BibitemShut {NoStop}%
\bibitem [{\citenamefont {Staub}(2008)}]{Staub:2008uz}%
  \BibitemOpen
  \bibfield  {author} {\bibinfo {author} {\bibfnamefont {F.}~\bibnamefont {Staub}},\ }\href@noop {} {\  (\bibinfo {year} {2008})},\ \Eprint {http://arxiv.org/abs/0806.0538} {arXiv:0806.0538 [hep-ph]} \BibitemShut {NoStop}%
\bibitem [{\citenamefont {Porod}(2003)}]{Porod:2003um}%
  \BibitemOpen
  \bibfield  {author} {\bibinfo {author} {\bibfnamefont {W.}~\bibnamefont {Porod}},\ }\href {\doibase 10.1016/S0010-4655(03)00222-4} {\bibfield  {journal} {\bibinfo  {journal} {Comput. Phys. Commun.}\ }\textbf {\bibinfo {volume} {153}},\ \bibinfo {pages} {275} (\bibinfo {year} {2003})},\ \Eprint {http://arxiv.org/abs/hep-ph/0301101} {arXiv:hep-ph/0301101} \BibitemShut {NoStop}%
\bibitem [{\citenamefont {Porod}\ and\ \citenamefont {Staub}(2012)}]{Porod:2011nf}%
  \BibitemOpen
  \bibfield  {author} {\bibinfo {author} {\bibfnamefont {W.}~\bibnamefont {Porod}}\ and\ \bibinfo {author} {\bibfnamefont {F.}~\bibnamefont {Staub}},\ }\href {\doibase 10.1016/j.cpc.2012.05.021} {\bibfield  {journal} {\bibinfo  {journal} {Comput. Phys. Commun.}\ }\textbf {\bibinfo {volume} {183}},\ \bibinfo {pages} {2458} (\bibinfo {year} {2012})},\ \Eprint {http://arxiv.org/abs/1104.1573} {arXiv:1104.1573 [hep-ph]} \BibitemShut {NoStop}%
\bibitem [{\citenamefont {Bahl}\ \emph {et~al.}(2023)\citenamefont {Bahl}, \citenamefont {Biek\"otter}, \citenamefont {Heinemeyer}, \citenamefont {Li}, \citenamefont {Paasch}, \citenamefont {Weiglein},\ and\ \citenamefont {Wittbrodt}}]{Bahl:2022igd}%
  \BibitemOpen
  \bibfield  {author} {\bibinfo {author} {\bibfnamefont {H.}~\bibnamefont {Bahl}}, \bibinfo {author} {\bibfnamefont {T.}~\bibnamefont {Biek\"otter}}, \bibinfo {author} {\bibfnamefont {S.}~\bibnamefont {Heinemeyer}}, \bibinfo {author} {\bibfnamefont {C.}~\bibnamefont {Li}}, \bibinfo {author} {\bibfnamefont {S.}~\bibnamefont {Paasch}}, \bibinfo {author} {\bibfnamefont {G.}~\bibnamefont {Weiglein}}, \ and\ \bibinfo {author} {\bibfnamefont {J.}~\bibnamefont {Wittbrodt}},\ }\href {\doibase 10.1016/j.cpc.2023.108803} {\bibfield  {journal} {\bibinfo  {journal} {Comput. Phys. Commun.}\ }\textbf {\bibinfo {volume} {291}},\ \bibinfo {pages} {108803} (\bibinfo {year} {2023})},\ \Eprint {http://arxiv.org/abs/2210.09332} {arXiv:2210.09332 [hep-ph]} \BibitemShut {NoStop}%
\bibitem [{\citenamefont {Bechtle}\ \emph {et~al.}(2010)\citenamefont {Bechtle}, \citenamefont {Brein}, \citenamefont {Heinemeyer}, \citenamefont {Weiglein},\ and\ \citenamefont {Williams}}]{Bechtle:2008jh}%
  \BibitemOpen
  \bibfield  {author} {\bibinfo {author} {\bibfnamefont {P.}~\bibnamefont {Bechtle}}, \bibinfo {author} {\bibfnamefont {O.}~\bibnamefont {Brein}}, \bibinfo {author} {\bibfnamefont {S.}~\bibnamefont {Heinemeyer}}, \bibinfo {author} {\bibfnamefont {G.}~\bibnamefont {Weiglein}}, \ and\ \bibinfo {author} {\bibfnamefont {K.~E.}\ \bibnamefont {Williams}},\ }\href {\doibase 10.1016/j.cpc.2009.09.003} {\bibfield  {journal} {\bibinfo  {journal} {Comput. Phys. Commun.}\ }\textbf {\bibinfo {volume} {181}},\ \bibinfo {pages} {138} (\bibinfo {year} {2010})},\ \Eprint {http://arxiv.org/abs/0811.4169} {arXiv:0811.4169 [hep-ph]} \BibitemShut {NoStop}%
\bibitem [{\citenamefont {Bechtle}\ \emph {et~al.}(2011)\citenamefont {Bechtle}, \citenamefont {Brein}, \citenamefont {Heinemeyer}, \citenamefont {Weiglein},\ and\ \citenamefont {Williams}}]{Bechtle:2011sb}%
  \BibitemOpen
  \bibfield  {author} {\bibinfo {author} {\bibfnamefont {P.}~\bibnamefont {Bechtle}}, \bibinfo {author} {\bibfnamefont {O.}~\bibnamefont {Brein}}, \bibinfo {author} {\bibfnamefont {S.}~\bibnamefont {Heinemeyer}}, \bibinfo {author} {\bibfnamefont {G.}~\bibnamefont {Weiglein}}, \ and\ \bibinfo {author} {\bibfnamefont {K.~E.}\ \bibnamefont {Williams}},\ }\href {\doibase 10.1016/j.cpc.2011.07.015} {\bibfield  {journal} {\bibinfo  {journal} {Comput. Phys. Commun.}\ }\textbf {\bibinfo {volume} {182}},\ \bibinfo {pages} {2605} (\bibinfo {year} {2011})},\ \Eprint {http://arxiv.org/abs/1102.1898} {arXiv:1102.1898 [hep-ph]} \BibitemShut {NoStop}%
\bibitem [{\citenamefont {Bechtle}\ \emph {et~al.}(2012)\citenamefont {Bechtle}, \citenamefont {Brein}, \citenamefont {Heinemeyer}, \citenamefont {Stal}, \citenamefont {Stefaniak}, \citenamefont {Weiglein},\ and\ \citenamefont {Williams}}]{Bechtle:2012lvg}%
  \BibitemOpen
  \bibfield  {author} {\bibinfo {author} {\bibfnamefont {P.}~\bibnamefont {Bechtle}}, \bibinfo {author} {\bibfnamefont {O.}~\bibnamefont {Brein}}, \bibinfo {author} {\bibfnamefont {S.}~\bibnamefont {Heinemeyer}}, \bibinfo {author} {\bibfnamefont {O.}~\bibnamefont {Stal}}, \bibinfo {author} {\bibfnamefont {T.}~\bibnamefont {Stefaniak}}, \bibinfo {author} {\bibfnamefont {G.}~\bibnamefont {Weiglein}}, \ and\ \bibinfo {author} {\bibfnamefont {K.}~\bibnamefont {Williams}},\ }\href {\doibase 10.22323/1.156.0024} {\bibfield  {journal} {\bibinfo  {journal} {PoS}\ }\textbf {\bibinfo {volume} {CHARGED2012}},\ \bibinfo {pages} {024} (\bibinfo {year} {2012})},\ \Eprint {http://arxiv.org/abs/1301.2345} {arXiv:1301.2345 [hep-ph]} \BibitemShut {NoStop}%
\bibitem [{\citenamefont {Bechtle}\ \emph {et~al.}(2014{\natexlab{a}})\citenamefont {Bechtle}, \citenamefont {Brein}, \citenamefont {Heinemeyer}, \citenamefont {St\r{a}l}, \citenamefont {Stefaniak}, \citenamefont {Weiglein},\ and\ \citenamefont {Williams}}]{Bechtle:2013wla}%
  \BibitemOpen
  \bibfield  {author} {\bibinfo {author} {\bibfnamefont {P.}~\bibnamefont {Bechtle}}, \bibinfo {author} {\bibfnamefont {O.}~\bibnamefont {Brein}}, \bibinfo {author} {\bibfnamefont {S.}~\bibnamefont {Heinemeyer}}, \bibinfo {author} {\bibfnamefont {O.}~\bibnamefont {St\r{a}l}}, \bibinfo {author} {\bibfnamefont {T.}~\bibnamefont {Stefaniak}}, \bibinfo {author} {\bibfnamefont {G.}~\bibnamefont {Weiglein}}, \ and\ \bibinfo {author} {\bibfnamefont {K.~E.}\ \bibnamefont {Williams}},\ }\href {\doibase 10.1140/epjc/s10052-013-2693-2} {\bibfield  {journal} {\bibinfo  {journal} {Eur. Phys. J. C}\ }\textbf {\bibinfo {volume} {74}},\ \bibinfo {pages} {2693} (\bibinfo {year} {2014}{\natexlab{a}})},\ \Eprint {http://arxiv.org/abs/1311.0055} {arXiv:1311.0055 [hep-ph]} \BibitemShut {NoStop}%
\bibitem [{\citenamefont {Bechtle}\ \emph {et~al.}(2015)\citenamefont {Bechtle}, \citenamefont {Heinemeyer}, \citenamefont {Stal}, \citenamefont {Stefaniak},\ and\ \citenamefont {Weiglein}}]{Bechtle:2015pma}%
  \BibitemOpen
  \bibfield  {author} {\bibinfo {author} {\bibfnamefont {P.}~\bibnamefont {Bechtle}}, \bibinfo {author} {\bibfnamefont {S.}~\bibnamefont {Heinemeyer}}, \bibinfo {author} {\bibfnamefont {O.}~\bibnamefont {Stal}}, \bibinfo {author} {\bibfnamefont {T.}~\bibnamefont {Stefaniak}}, \ and\ \bibinfo {author} {\bibfnamefont {G.}~\bibnamefont {Weiglein}},\ }\href {\doibase 10.1140/epjc/s10052-015-3650-z} {\bibfield  {journal} {\bibinfo  {journal} {Eur. Phys. J. C}\ }\textbf {\bibinfo {volume} {75}},\ \bibinfo {pages} {421} (\bibinfo {year} {2015})},\ \Eprint {http://arxiv.org/abs/1507.06706} {arXiv:1507.06706 [hep-ph]} \BibitemShut {NoStop}%
\bibitem [{\citenamefont {Bechtle}\ \emph {et~al.}(2020)\citenamefont {Bechtle}, \citenamefont {Dercks}, \citenamefont {Heinemeyer}, \citenamefont {Klingl}, \citenamefont {Stefaniak}, \citenamefont {Weiglein},\ and\ \citenamefont {Wittbrodt}}]{Bechtle:2020pkv}%
  \BibitemOpen
  \bibfield  {author} {\bibinfo {author} {\bibfnamefont {P.}~\bibnamefont {Bechtle}}, \bibinfo {author} {\bibfnamefont {D.}~\bibnamefont {Dercks}}, \bibinfo {author} {\bibfnamefont {S.}~\bibnamefont {Heinemeyer}}, \bibinfo {author} {\bibfnamefont {T.}~\bibnamefont {Klingl}}, \bibinfo {author} {\bibfnamefont {T.}~\bibnamefont {Stefaniak}}, \bibinfo {author} {\bibfnamefont {G.}~\bibnamefont {Weiglein}}, \ and\ \bibinfo {author} {\bibfnamefont {J.}~\bibnamefont {Wittbrodt}},\ }\href {\doibase 10.1140/epjc/s10052-020-08557-9} {\bibfield  {journal} {\bibinfo  {journal} {Eur. Phys. J. C}\ }\textbf {\bibinfo {volume} {80}},\ \bibinfo {pages} {1211} (\bibinfo {year} {2020})},\ \Eprint {http://arxiv.org/abs/2006.06007} {arXiv:2006.06007 [hep-ph]} \BibitemShut {NoStop}%
\bibitem [{\citenamefont {Bahl}\ \emph {et~al.}(2022)\citenamefont {Bahl}, \citenamefont {Lozano}, \citenamefont {Stefaniak},\ and\ \citenamefont {Wittbrodt}}]{Bahl:2021yhk}%
  \BibitemOpen
  \bibfield  {author} {\bibinfo {author} {\bibfnamefont {H.}~\bibnamefont {Bahl}}, \bibinfo {author} {\bibfnamefont {V.~M.}\ \bibnamefont {Lozano}}, \bibinfo {author} {\bibfnamefont {T.}~\bibnamefont {Stefaniak}}, \ and\ \bibinfo {author} {\bibfnamefont {J.}~\bibnamefont {Wittbrodt}},\ }\href {\doibase 10.1140/epjc/s10052-022-10446-2} {\bibfield  {journal} {\bibinfo  {journal} {Eur. Phys. J. C}\ }\textbf {\bibinfo {volume} {82}},\ \bibinfo {pages} {584} (\bibinfo {year} {2022})},\ \Eprint {http://arxiv.org/abs/2109.10366} {arXiv:2109.10366 [hep-ph]} \BibitemShut {NoStop}%
\bibitem [{\citenamefont {Bechtle}\ \emph {et~al.}(2014{\natexlab{b}})\citenamefont {Bechtle}, \citenamefont {Heinemeyer}, \citenamefont {St\r{a}l}, \citenamefont {Stefaniak},\ and\ \citenamefont {Weiglein}}]{Bechtle:2013xfa}%
  \BibitemOpen
  \bibfield  {author} {\bibinfo {author} {\bibfnamefont {P.}~\bibnamefont {Bechtle}}, \bibinfo {author} {\bibfnamefont {S.}~\bibnamefont {Heinemeyer}}, \bibinfo {author} {\bibfnamefont {O.}~\bibnamefont {St\r{a}l}}, \bibinfo {author} {\bibfnamefont {T.}~\bibnamefont {Stefaniak}}, \ and\ \bibinfo {author} {\bibfnamefont {G.}~\bibnamefont {Weiglein}},\ }\href {\doibase 10.1140/epjc/s10052-013-2711-4} {\bibfield  {journal} {\bibinfo  {journal} {Eur. Phys. J. C}\ }\textbf {\bibinfo {volume} {74}},\ \bibinfo {pages} {2711} (\bibinfo {year} {2014}{\natexlab{b}})},\ \Eprint {http://arxiv.org/abs/1305.1933} {arXiv:1305.1933 [hep-ph]} \BibitemShut {NoStop}%
\bibitem [{\citenamefont {St\r{a}l}\ and\ \citenamefont {Stefaniak}(2013)}]{Stal:2013hwa}%
  \BibitemOpen
  \bibfield  {author} {\bibinfo {author} {\bibfnamefont {O.}~\bibnamefont {St\r{a}l}}\ and\ \bibinfo {author} {\bibfnamefont {T.}~\bibnamefont {Stefaniak}},\ }\href {\doibase 10.22323/1.180.0314} {\bibfield  {journal} {\bibinfo  {journal} {PoS}\ }\textbf {\bibinfo {volume} {EPS-HEP2013}},\ \bibinfo {pages} {314} (\bibinfo {year} {2013})},\ \Eprint {http://arxiv.org/abs/1310.4039} {arXiv:1310.4039 [hep-ph]} \BibitemShut {NoStop}%
\bibitem [{\citenamefont {Bechtle}\ \emph {et~al.}(2014{\natexlab{c}})\citenamefont {Bechtle}, \citenamefont {Heinemeyer}, \citenamefont {St\r{a}l}, \citenamefont {Stefaniak},\ and\ \citenamefont {Weiglein}}]{Bechtle:2014ewa}%
  \BibitemOpen
  \bibfield  {author} {\bibinfo {author} {\bibfnamefont {P.}~\bibnamefont {Bechtle}}, \bibinfo {author} {\bibfnamefont {S.}~\bibnamefont {Heinemeyer}}, \bibinfo {author} {\bibfnamefont {O.}~\bibnamefont {St\r{a}l}}, \bibinfo {author} {\bibfnamefont {T.}~\bibnamefont {Stefaniak}}, \ and\ \bibinfo {author} {\bibfnamefont {G.}~\bibnamefont {Weiglein}},\ }\href {\doibase 10.1007/JHEP11(2014)039} {\bibfield  {journal} {\bibinfo  {journal} {JHEP}\ }\textbf {\bibinfo {volume} {11}},\ \bibinfo {pages} {039} (\bibinfo {year} {2014}{\natexlab{c}})},\ \Eprint {http://arxiv.org/abs/1403.1582} {arXiv:1403.1582 [hep-ph]} \BibitemShut {NoStop}%
\bibitem [{\citenamefont {Bechtle}\ \emph {et~al.}(2021)\citenamefont {Bechtle}, \citenamefont {Heinemeyer}, \citenamefont {Klingl}, \citenamefont {Stefaniak}, \citenamefont {Weiglein},\ and\ \citenamefont {Wittbrodt}}]{Bechtle:2020uwn}%
  \BibitemOpen
  \bibfield  {author} {\bibinfo {author} {\bibfnamefont {P.}~\bibnamefont {Bechtle}}, \bibinfo {author} {\bibfnamefont {S.}~\bibnamefont {Heinemeyer}}, \bibinfo {author} {\bibfnamefont {T.}~\bibnamefont {Klingl}}, \bibinfo {author} {\bibfnamefont {T.}~\bibnamefont {Stefaniak}}, \bibinfo {author} {\bibfnamefont {G.}~\bibnamefont {Weiglein}}, \ and\ \bibinfo {author} {\bibfnamefont {J.}~\bibnamefont {Wittbrodt}},\ }\href {\doibase 10.1140/epjc/s10052-021-08942-y} {\bibfield  {journal} {\bibinfo  {journal} {Eur. Phys. J. C}\ }\textbf {\bibinfo {volume} {81}},\ \bibinfo {pages} {145} (\bibinfo {year} {2021})},\ \Eprint {http://arxiv.org/abs/2012.09197} {arXiv:2012.09197 [hep-ph]} \BibitemShut {NoStop}%
\bibitem [{\citenamefont {Belanger}\ \emph {et~al.}(2007)\citenamefont {Belanger}, \citenamefont {Boudjema}, \citenamefont {Pukhov},\ and\ \citenamefont {Semenov}}]{Belanger:2006is}%
  \BibitemOpen
  \bibfield  {author} {\bibinfo {author} {\bibfnamefont {G.}~\bibnamefont {Belanger}}, \bibinfo {author} {\bibfnamefont {F.}~\bibnamefont {Boudjema}}, \bibinfo {author} {\bibfnamefont {A.}~\bibnamefont {Pukhov}}, \ and\ \bibinfo {author} {\bibfnamefont {A.}~\bibnamefont {Semenov}},\ }\href {\doibase 10.1016/j.cpc.2006.11.008} {\bibfield  {journal} {\bibinfo  {journal} {Comput. Phys. Commun.}\ }\textbf {\bibinfo {volume} {176}},\ \bibinfo {pages} {367} (\bibinfo {year} {2007})},\ \Eprint {http://arxiv.org/abs/hep-ph/0607059} {arXiv:hep-ph/0607059} \BibitemShut {NoStop}%
\bibitem [{\citenamefont {Belanger}\ \emph {et~al.}(2009)\citenamefont {Belanger}, \citenamefont {Boudjema}, \citenamefont {Pukhov},\ and\ \citenamefont {Semenov}}]{Belanger:2008sj}%
  \BibitemOpen
  \bibfield  {author} {\bibinfo {author} {\bibfnamefont {G.}~\bibnamefont {Belanger}}, \bibinfo {author} {\bibfnamefont {F.}~\bibnamefont {Boudjema}}, \bibinfo {author} {\bibfnamefont {A.}~\bibnamefont {Pukhov}}, \ and\ \bibinfo {author} {\bibfnamefont {A.}~\bibnamefont {Semenov}},\ }\href {\doibase 10.1016/j.cpc.2008.11.019} {\bibfield  {journal} {\bibinfo  {journal} {Comput. Phys. Commun.}\ }\textbf {\bibinfo {volume} {180}},\ \bibinfo {pages} {747} (\bibinfo {year} {2009})},\ \Eprint {http://arxiv.org/abs/0803.2360} {arXiv:0803.2360 [hep-ph]} \BibitemShut {NoStop}%
\bibitem [{\citenamefont {Belanger}\ \emph {et~al.}(2010)\citenamefont {Belanger}, \citenamefont {Boudjema}, \citenamefont {Pukhov},\ and\ \citenamefont {Semenov}}]{Belanger:2010pz}%
  \BibitemOpen
  \bibfield  {author} {\bibinfo {author} {\bibfnamefont {G.}~\bibnamefont {Belanger}}, \bibinfo {author} {\bibfnamefont {F.}~\bibnamefont {Boudjema}}, \bibinfo {author} {\bibfnamefont {A.}~\bibnamefont {Pukhov}}, \ and\ \bibinfo {author} {\bibfnamefont {A.}~\bibnamefont {Semenov}},\ }\href {\doibase 10.1393/ncc/i2010-10591-3} {\bibfield  {journal} {\bibinfo  {journal} {Nuovo Cim. C}\ }\textbf {\bibinfo {volume} {033N2}},\ \bibinfo {pages} {111} (\bibinfo {year} {2010})},\ \Eprint {http://arxiv.org/abs/1005.4133} {arXiv:1005.4133 [hep-ph]} \BibitemShut {NoStop}%
\bibitem [{\citenamefont {Belanger}\ \emph {et~al.}(2014)\citenamefont {Belanger}, \citenamefont {Boudjema}, \citenamefont {Pukhov},\ and\ \citenamefont {Semenov}}]{Belanger:2013oya}%
  \BibitemOpen
  \bibfield  {author} {\bibinfo {author} {\bibfnamefont {G.}~\bibnamefont {Belanger}}, \bibinfo {author} {\bibfnamefont {F.}~\bibnamefont {Boudjema}}, \bibinfo {author} {\bibfnamefont {A.}~\bibnamefont {Pukhov}}, \ and\ \bibinfo {author} {\bibfnamefont {A.}~\bibnamefont {Semenov}},\ }\href {\doibase 10.1016/j.cpc.2013.10.016} {\bibfield  {journal} {\bibinfo  {journal} {Comput. Phys. Commun.}\ }\textbf {\bibinfo {volume} {185}},\ \bibinfo {pages} {960} (\bibinfo {year} {2014})},\ \Eprint {http://arxiv.org/abs/1305.0237} {arXiv:1305.0237 [hep-ph]} \BibitemShut {NoStop}%
\bibitem [{\citenamefont {B\'elanger}\ \emph {et~al.}(2015)\citenamefont {B\'elanger}, \citenamefont {Boudjema}, \citenamefont {Pukhov},\ and\ \citenamefont {Semenov}}]{Belanger:2014vza}%
  \BibitemOpen
  \bibfield  {author} {\bibinfo {author} {\bibfnamefont {G.}~\bibnamefont {B\'elanger}}, \bibinfo {author} {\bibfnamefont {F.}~\bibnamefont {Boudjema}}, \bibinfo {author} {\bibfnamefont {A.}~\bibnamefont {Pukhov}}, \ and\ \bibinfo {author} {\bibfnamefont {A.}~\bibnamefont {Semenov}},\ }\href {\doibase 10.1016/j.cpc.2015.03.003} {\bibfield  {journal} {\bibinfo  {journal} {Comput. Phys. Commun.}\ }\textbf {\bibinfo {volume} {192}},\ \bibinfo {pages} {322} (\bibinfo {year} {2015})},\ \Eprint {http://arxiv.org/abs/1407.6129} {arXiv:1407.6129 [hep-ph]} \BibitemShut {NoStop}%
\bibitem [{\citenamefont {B\'elanger}\ \emph {et~al.}(2018)\citenamefont {B\'elanger}, \citenamefont {Boudjema}, \citenamefont {Goudelis}, \citenamefont {Pukhov},\ and\ \citenamefont {Zaldivar}}]{Belanger:2018ccd}%
  \BibitemOpen
  \bibfield  {author} {\bibinfo {author} {\bibfnamefont {G.}~\bibnamefont {B\'elanger}}, \bibinfo {author} {\bibfnamefont {F.}~\bibnamefont {Boudjema}}, \bibinfo {author} {\bibfnamefont {A.}~\bibnamefont {Goudelis}}, \bibinfo {author} {\bibfnamefont {A.}~\bibnamefont {Pukhov}}, \ and\ \bibinfo {author} {\bibfnamefont {B.}~\bibnamefont {Zaldivar}},\ }\href {\doibase 10.1016/j.cpc.2018.04.027} {\bibfield  {journal} {\bibinfo  {journal} {Comput. Phys. Commun.}\ }\textbf {\bibinfo {volume} {231}},\ \bibinfo {pages} {173} (\bibinfo {year} {2018})},\ \Eprint {http://arxiv.org/abs/1801.03509} {arXiv:1801.03509 [hep-ph]} \BibitemShut {NoStop}%
\bibitem [{\citenamefont {Altakach}\ \emph {et~al.}(2024)\citenamefont {Altakach}, \citenamefont {Kraml}, \citenamefont {Lessa}, \citenamefont {Narasimha}, \citenamefont {Pascal}, \citenamefont {Ramos}, \citenamefont {Villamizar},\ and\ \citenamefont {Waltenberger}}]{smodels:v3}%
  \BibitemOpen
  \bibfield  {author} {\bibinfo {author} {\bibfnamefont {M.~M.}\ \bibnamefont {Altakach}}, \bibinfo {author} {\bibfnamefont {S.}~\bibnamefont {Kraml}}, \bibinfo {author} {\bibfnamefont {A.}~\bibnamefont {Lessa}}, \bibinfo {author} {\bibfnamefont {S.}~\bibnamefont {Narasimha}}, \bibinfo {author} {\bibfnamefont {T.}~\bibnamefont {Pascal}}, \bibinfo {author} {\bibfnamefont {C.}~\bibnamefont {Ramos}}, \bibinfo {author} {\bibfnamefont {Y.}~\bibnamefont {Villamizar}}, \ and\ \bibinfo {author} {\bibfnamefont {W.}~\bibnamefont {Waltenberger}},\ }\href@noop {} {\  (\bibinfo {year} {2024})},\ \Eprint {http://arxiv.org/abs/240x.nnnnn} {arXiv:240x.nnnnn [hep-ph]} \BibitemShut {NoStop}%
\bibitem [{\citenamefont {Goodsell}\ and\ \citenamefont {Joury}(2024)}]{Goodsell:2023iac}%
  \BibitemOpen
  \bibfield  {author} {\bibinfo {author} {\bibfnamefont {M.~D.}\ \bibnamefont {Goodsell}}\ and\ \bibinfo {author} {\bibfnamefont {A.}~\bibnamefont {Joury}},\ }\href {\doibase 10.1016/j.cpc.2023.109057} {\bibfield  {journal} {\bibinfo  {journal} {Comput. Phys. Commun.}\ }\textbf {\bibinfo {volume} {297}},\ \bibinfo {pages} {109057} (\bibinfo {year} {2024})},\ \Eprint {http://arxiv.org/abs/2301.01154} {arXiv:2301.01154 [hep-ph]} \BibitemShut {NoStop}%
\bibitem [{\citenamefont {Faraggi}\ and\ \citenamefont {Goodsell}(2024)}]{Faraggi:2023jzm}%
  \BibitemOpen
  \bibfield  {author} {\bibinfo {author} {\bibfnamefont {A.~E.}\ \bibnamefont {Faraggi}}\ and\ \bibinfo {author} {\bibfnamefont {M.~D.}\ \bibnamefont {Goodsell}},\ }\href {\doibase 10.1140/epjc/s10052-024-12900-9} {\bibfield  {journal} {\bibinfo  {journal} {Eur. Phys. J. C}\ }\textbf {\bibinfo {volume} {84}},\ \bibinfo {pages} {589} (\bibinfo {year} {2024})},\ \Eprint {http://arxiv.org/abs/2312.13411} {arXiv:2312.13411 [hep-ph]} \BibitemShut {NoStop}%
\bibitem [{\citenamefont {Fuks}\ \emph {et~al.}(2025)\citenamefont {Fuks}, \citenamefont {Goodsell},\ and\ \citenamefont {Murphy}}]{Fuks:2024qdt}%
  \BibitemOpen
  \bibfield  {author} {\bibinfo {author} {\bibfnamefont {B.}~\bibnamefont {Fuks}}, \bibinfo {author} {\bibfnamefont {M.~D.}\ \bibnamefont {Goodsell}}, \ and\ \bibinfo {author} {\bibfnamefont {T.}~\bibnamefont {Murphy}},\ }\href {\doibase 10.1103/PhysRevD.111.055010} {\bibfield  {journal} {\bibinfo  {journal} {Phys. Rev. D}\ }\textbf {\bibinfo {volume} {111}},\ \bibinfo {pages} {055010} (\bibinfo {year} {2025})},\ \Eprint {http://arxiv.org/abs/2409.03014} {arXiv:2409.03014 [hep-ph]} \BibitemShut {NoStop}%
\bibitem [{\citenamefont {Agin}\ \emph {et~al.}(2024{\natexlab{a}})\citenamefont {Agin}, \citenamefont {Fuks}, \citenamefont {Goodsell},\ and\ \citenamefont {Murphy}}]{Agin:2023yoq}%
  \BibitemOpen
  \bibfield  {author} {\bibinfo {author} {\bibfnamefont {D.}~\bibnamefont {Agin}}, \bibinfo {author} {\bibfnamefont {B.}~\bibnamefont {Fuks}}, \bibinfo {author} {\bibfnamefont {M.~D.}\ \bibnamefont {Goodsell}}, \ and\ \bibinfo {author} {\bibfnamefont {T.}~\bibnamefont {Murphy}},\ }\href {\doibase 10.1016/j.physletb.2024.138597} {\bibfield  {journal} {\bibinfo  {journal} {Phys. Lett. B}\ }\textbf {\bibinfo {volume} {853}},\ \bibinfo {pages} {138597} (\bibinfo {year} {2024}{\natexlab{a}})},\ \Eprint {http://arxiv.org/abs/2311.17149} {arXiv:2311.17149 [hep-ph]} \BibitemShut {NoStop}%
\bibitem [{\citenamefont {Agin}\ \emph {et~al.}(2024{\natexlab{b}})\citenamefont {Agin}, \citenamefont {Fuks}, \citenamefont {Goodsell},\ and\ \citenamefont {Murphy}}]{Agin:2024yfs}%
  \BibitemOpen
  \bibfield  {author} {\bibinfo {author} {\bibfnamefont {D.}~\bibnamefont {Agin}}, \bibinfo {author} {\bibfnamefont {B.}~\bibnamefont {Fuks}}, \bibinfo {author} {\bibfnamefont {M.~D.}\ \bibnamefont {Goodsell}}, \ and\ \bibinfo {author} {\bibfnamefont {T.}~\bibnamefont {Murphy}},\ }\href {\doibase 10.1140/epjc/s10052-024-13594-9} {\bibfield  {journal} {\bibinfo  {journal} {Eur. Phys. J. C}\ }\textbf {\bibinfo {volume} {84}},\ \bibinfo {pages} {1218} (\bibinfo {year} {2024}{\natexlab{b}})},\ \Eprint {http://arxiv.org/abs/2404.12423} {arXiv:2404.12423 [hep-ph]} \BibitemShut {NoStop}%
\bibitem [{\citenamefont {Benakli}\ \emph {et~al.}(2023)\citenamefont {Benakli}, \citenamefont {Goodsell}, \citenamefont {Ke},\ and\ \citenamefont {Slavich}}]{Benakli:2022gjn}%
  \BibitemOpen
  \bibfield  {author} {\bibinfo {author} {\bibfnamefont {K.}~\bibnamefont {Benakli}}, \bibinfo {author} {\bibfnamefont {M.}~\bibnamefont {Goodsell}}, \bibinfo {author} {\bibfnamefont {W.}~\bibnamefont {Ke}}, \ and\ \bibinfo {author} {\bibfnamefont {P.}~\bibnamefont {Slavich}},\ }\href {\doibase 10.1140/epjc/s10052-022-11132-z} {\bibfield  {journal} {\bibinfo  {journal} {Eur. Phys. J. C}\ }\textbf {\bibinfo {volume} {83}},\ \bibinfo {pages} {43} (\bibinfo {year} {2023})},\ \Eprint {http://arxiv.org/abs/2208.05867} {arXiv:2208.05867 [hep-ph]} \BibitemShut {NoStop}%
\bibitem [{\citenamefont {Goodsell}\ and\ \citenamefont {Joury}(2023)}]{Goodsell:2022beo}%
  \BibitemOpen
  \bibfield  {author} {\bibinfo {author} {\bibfnamefont {M.~D.}\ \bibnamefont {Goodsell}}\ and\ \bibinfo {author} {\bibfnamefont {A.}~\bibnamefont {Joury}},\ }\href {\doibase 10.1140/epjc/s10052-023-11368-3} {\bibfield  {journal} {\bibinfo  {journal} {Eur. Phys. J. C}\ }\textbf {\bibinfo {volume} {83}},\ \bibinfo {pages} {268} (\bibinfo {year} {2023})},\ \Eprint {http://arxiv.org/abs/2204.13950} {arXiv:2204.13950 [hep-ph]} \BibitemShut {NoStop}%
\bibitem [{\citenamefont {Goodsell}\ and\ \citenamefont {Priya}(2022)}]{Goodsell:2021iwc}%
  \BibitemOpen
  \bibfield  {author} {\bibinfo {author} {\bibfnamefont {M.~D.}\ \bibnamefont {Goodsell}}\ and\ \bibinfo {author} {\bibfnamefont {L.}~\bibnamefont {Priya}},\ }\href {\doibase 10.1140/epjc/s10052-022-10188-1} {\bibfield  {journal} {\bibinfo  {journal} {Eur. Phys. J. C}\ }\textbf {\bibinfo {volume} {82}},\ \bibinfo {pages} {235} (\bibinfo {year} {2022})},\ \Eprint {http://arxiv.org/abs/2106.08815} {arXiv:2106.08815 [hep-ph]} \BibitemShut {NoStop}%
\bibitem [{\citenamefont {Goodsell}\ and\ \citenamefont {Moutafis}(2021)}]{Goodsell:2020rfu}%
  \BibitemOpen
  \bibfield  {author} {\bibinfo {author} {\bibfnamefont {M.~D.}\ \bibnamefont {Goodsell}}\ and\ \bibinfo {author} {\bibfnamefont {R.}~\bibnamefont {Moutafis}},\ }\href {\doibase 10.1140/epjc/s10052-021-09597-5} {\bibfield  {journal} {\bibinfo  {journal} {Eur. Phys. J. C}\ }\textbf {\bibinfo {volume} {81}},\ \bibinfo {pages} {808} (\bibinfo {year} {2021})},\ \Eprint {http://arxiv.org/abs/2012.09022} {arXiv:2012.09022 [hep-ph]} \BibitemShut {NoStop}%
\bibitem [{\citenamefont {Dom\`enech}\ \emph {et~al.}(2021)\citenamefont {Dom\`enech}, \citenamefont {Goodsell},\ and\ \citenamefont {Wetterich}}]{Domenech:2020yjf}%
  \BibitemOpen
  \bibfield  {author} {\bibinfo {author} {\bibfnamefont {G.}~\bibnamefont {Dom\`enech}}, \bibinfo {author} {\bibfnamefont {M.}~\bibnamefont {Goodsell}}, \ and\ \bibinfo {author} {\bibfnamefont {C.}~\bibnamefont {Wetterich}},\ }\href {\doibase 10.1007/JHEP01(2021)180} {\bibfield  {journal} {\bibinfo  {journal} {JHEP}\ }\textbf {\bibinfo {volume} {01}},\ \bibinfo {pages} {180} (\bibinfo {year} {2021})},\ \Eprint {http://arxiv.org/abs/2008.04310} {arXiv:2008.04310 [hep-ph]} \BibitemShut {NoStop}%
\bibitem [{\citenamefont {Feroz}\ and\ \citenamefont {Hobson}(2008)}]{Feroz:2007kg}%
  \BibitemOpen
  \bibfield  {author} {\bibinfo {author} {\bibfnamefont {F.}~\bibnamefont {Feroz}}\ and\ \bibinfo {author} {\bibfnamefont {M.~P.}\ \bibnamefont {Hobson}},\ }\href {\doibase 10.1111/j.1365-2966.2007.12353.x} {\bibfield  {journal} {\bibinfo  {journal} {Mon. Not. Roy. Astron. Soc.}\ }\textbf {\bibinfo {volume} {384}},\ \bibinfo {pages} {449} (\bibinfo {year} {2008})},\ \Eprint {http://arxiv.org/abs/0704.3704} {arXiv:0704.3704 [astro-ph]} \BibitemShut {NoStop}%
\bibitem [{\citenamefont {Feroz}\ \emph {et~al.}(2009)\citenamefont {Feroz}, \citenamefont {Hobson},\ and\ \citenamefont {Bridges}}]{Feroz:2008xx}%
  \BibitemOpen
  \bibfield  {author} {\bibinfo {author} {\bibfnamefont {F.}~\bibnamefont {Feroz}}, \bibinfo {author} {\bibfnamefont {M.~P.}\ \bibnamefont {Hobson}}, \ and\ \bibinfo {author} {\bibfnamefont {M.}~\bibnamefont {Bridges}},\ }\href {\doibase 10.1111/j.1365-2966.2009.14548.x} {\bibfield  {journal} {\bibinfo  {journal} {Mon. Not. Roy. Astron. Soc.}\ }\textbf {\bibinfo {volume} {398}},\ \bibinfo {pages} {1601} (\bibinfo {year} {2009})},\ \Eprint {http://arxiv.org/abs/0809.3437} {arXiv:0809.3437 [astro-ph]} \BibitemShut {NoStop}%
\bibitem [{\citenamefont {Feroz}\ \emph {et~al.}(2019)\citenamefont {Feroz}, \citenamefont {Hobson}, \citenamefont {Cameron},\ and\ \citenamefont {Pettitt}}]{Feroz:2013hea}%
  \BibitemOpen
  \bibfield  {author} {\bibinfo {author} {\bibfnamefont {F.}~\bibnamefont {Feroz}}, \bibinfo {author} {\bibfnamefont {M.~P.}\ \bibnamefont {Hobson}}, \bibinfo {author} {\bibfnamefont {E.}~\bibnamefont {Cameron}}, \ and\ \bibinfo {author} {\bibfnamefont {A.~N.}\ \bibnamefont {Pettitt}},\ }\href {\doibase 10.21105/astro.1306.2144} {\bibfield  {journal} {\bibinfo  {journal} {Open J. Astrophys.}\ }\textbf {\bibinfo {volume} {2}},\ \bibinfo {pages} {10} (\bibinfo {year} {2019})},\ \Eprint {http://arxiv.org/abs/1306.2144} {arXiv:1306.2144 [astro-ph.IM]} \BibitemShut {NoStop}%
\bibitem [{\citenamefont {Aghanim}\ \emph {et~al.}(2020)\citenamefont {Aghanim} \emph {et~al.}}]{Planck:2018vyg}%
  \BibitemOpen
  \bibfield  {author} {\bibinfo {author} {\bibfnamefont {N.}~\bibnamefont {Aghanim}} \emph {et~al.} (\bibinfo {collaboration} {Planck}),\ }\href {\doibase 10.1051/0004-6361/201833910} {\bibfield  {journal} {\bibinfo  {journal} {Astron. Astrophys.}\ }\textbf {\bibinfo {volume} {641}},\ \bibinfo {pages} {A6} (\bibinfo {year} {2020})},\ \bibinfo {note} {[Erratum: Astron.Astrophys. 652, C4 (2021)]},\ \Eprint {http://arxiv.org/abs/1807.06209} {arXiv:1807.06209 [astro-ph.CO]} \BibitemShut {NoStop}%
\end{thebibliography}%

\end{document}